\documentclass[10pt,letterpaper,twocolumn,prx,amsmath,amssymb,nobalancelastpage,aps,prx,showpacs,preprintnumbers,longbibliography]{revtex4-1}
\usepackage[latin9]{inputenc}
\usepackage{bm}
\usepackage{amsmath}
\usepackage{amssymb}
\usepackage{graphicx}
\makeatletter

\pdfpageheight\paperheight
\pdfpagewidth\paperwidth

\usepackage{dcolumn}% Align table columns on decimal point
\usepackage{bm}% bold math
\usepackage{slashed}
\usepackage{amsthm}
\usepackage{youngtab}
\usepackage{color}

\makeatother

\begin{document}

\title{Interplay between intrinsic and emergent topological protection on interacting helical modes}

\author{Raul A. Santos$^{1,2}$, D.B. Gutman$^{3}$ and Sam T. Carr$^{4}$}

\affiliation{$^{1}$T.C.M. Group, Cavendish Laboratory, University of Cambridge, J.J. Thomson Avenue, Cambridge, CB3 0HE, United Kingdom}
\affiliation{$^{2}$School of Physics \& Astronomy, University of Birmingham, Edgbaston, Birmingham, B15 2TT, United Kingdom}

\affiliation{$^{3}$Department of Physics, Bar-Ilan University, Ramat Gan, 52900,
Israel}

\affiliation{$^{4}$School of Physical Sciences, University of Kent, Canterbury
CT2 7NH, United Kingdom}
\begin{abstract}
The interplay between topology and interactions on the edge of a two
dimensional topological insulator with time reversal symmetry is studied.
We consider a simple non-interacting system of three helical channels with an inherent $\mathbb{Z}_{2}$ topological protection, and hence a zero-temperature conductance of 
$G=e^2/h$.  We show that when interactions are added to the model, the ground state exhibits two different phases as function of the interaction parameters. One of these 
phases is a trivial insulator at zero temperature, as the symmetry protecting the non-interacting topological phase is spontaneously broken.  
In this phase there is zero conductance ($G=0$) at zero-temperature.
The other phase displays enhanced topological properties, with a topologically protected zero-temperature conductance of $G=3e^2/h$ and an emergent 
$\mathbb{Z}_3$ symmetry not present in the lattice model. The neutral sector in this phase is described by a massive version of $\mathbb{Z}_{3}$ parafermions.  
This state is an example of a dynamically enhanced symmetry protected topological state.
\end{abstract}

\date{\today}

\maketitle
 
\section{Introduction}

Topology plays a central role in the modern understanding of different physical systems, ranging from superfluid Helium to elementary particles \cite{Thouless,Haldane,Volovik}.
In the context of solid state physics, one of the first phenomenon that was identified as having a topological origin was the integer quantum Hall effect (IQHE).
In the IQHE, the existence of protected chiral modes on the edge of the sample is a consequence of the existence of a non-trivial first
Chern number \cite{TKNN}.
The topological nature of these modes renders them robust against disorder and enforces 
a quantization of the conductance in units $e^{2}/h$, where $h$ is
the Planck constant and $e$ the electric charge. The inclusion of interactions
dramatically changes this picture, as occurs in the fractional quantum
Hall effect (FQHE), where the huge degeneracy between fractionally
filled many-body states is (partly) lifted by the interaction, creating a correlated
state with fractional conductance and exotic quasiparticles \cite{Tsui1982,Laughlin1983}.

In recent years, time-reversal (TR) invariant topological materials were
discovered, reviving the interest in topological systems. Examples of such topological insulators (TIs) are formed due to spin-orbit interaction
\cite{Kane2005,Bernevig2006,Bernevig2006a,Zhang08,Konig2008,Fu2007,Fu2007a,Fu2008,Hsieh2007,Hsieh2008,Hasan2010,Qi}
that is sufficiently strong to invert the $s$-like valence electronic states and 
$p$-like conduction electrons in different hetero-structures \cite{Konig2008,Roth2009}. In particular, in two spatial dimensions
non-interacting TIs display helical edge modes, and are characterized by a $\mathbb{Z}_{2}$ topological invariant,
which counts the parity of the number of edge modes. The electric conductance of a noninteracting TI is fixed as long as TR symmetry is preserved, due
to the destructive interference between the counterpropagating spin states around a nonmagnetic impurity.
The role of symmetry in these states is crucial to preserve the topological properties. It is for this reason that they are dubbed symmetry protected 
topological (SPT) states.

In general, for non-interacting disordered systems, the topological classification is fully
established \cite{Kitaev2009,Shinsei2010} and is uniquely determined by the symmetry class and dimensionality of the single particle Hamiltonian. 
Weak interactions can change the topological properties of a non-interacting system in different ways, e.g. by modifying the whole state including the bulk, or 
by changing the edge degrees of freedom in the system, without changing the overall structure in the bulk.
An example of the former corresponds to the interacting Kitaev chain \cite{Fidkowski2010}, where the inclusion of interactions allows to connect
adiabatically two Hamiltonians belonging to different non-interacting topological states, reducing the non-interacting classification of $\mathbb{Z}$ down to $\mathbb{Z}_{8}$.
On the other hand, when the characteristic interaction strength is smaller than the bulk gap energy, interactions can only induce a change at the edge degrees of freedom.
In this latter context, it has been recently found that the interactions may lead to an emergence of topologically non-trivial edge states, in systems that are topologically 
trivial on the bulk according to the non-interacting classification.

The simplest example of this kind of phenomena appears on the edge of a two dimensional TI supporting two parallel helical modes. Generically, in a non-interacting system, 
these modes can hybridise and be localised by the presence of sufficient density of impurities, making the system topologically trivial. Surprisingly, in the presence of 
interaction, there is some possibility for these modes to be protected against localisation, by a zero bias anomaly mechanism in the case of vanishing tunneling \cite{Santos2015} or by the emergence 
of an effective spin gap \cite{Santos2016,Kainaris2017} that suppresses single particle backscattering when tunneling is present. 
In these cases, the system displays topological signatures like a robust value of conductance, quantized in units of $e^{2}/h$ and fractionalised zero modes in domain wall 
configurations. This protection has also been predicted to appear in truly one dimensional systems with spin-orbit interaction \cite{Keselman2015, Kainaris2015}.

Another mechanism in which interactions can affect the topological properties of a non-trivial SPT state, is by inducing an spontaneous breaking of the protecting symmetry 
in the groundstate, rendering the state topologically trivial. Recently \cite{Kagalovsky2018}, it has been shown that in a general system of $N>2$ helical modes,
interactions can decrease the conductance of the system to zero at zero temperature by creating a groundstate that spontaneously breaks TR symmetry.

In this work, we focus on  a system of three coupled helical modes with inter-channel tunneling, corresponding to the edge structure of an integer TI. Because the number of 
modes is odd, this system is topologically nontrivial according to the non-interacting classification and disorder can at most localise two modes, leaving one helical mode 
free to carry the charge. We show that the interactions generate two distinct phases in which each of the effects discussed above can occur: in one phase, the intrinsic 
topology is destroyed through breaking of TR symmetry; while in the other phase, the intrinsic topological protection is enhanced through a new distinct emergent topological state, 
which protects all three helical modes against localisation.  Both of these states have a number of emergent energy scales with different characteristics, which we summarize 
below.

%%%%%%%%%%%%%%%%%%%%%%%%%%%%%%%%%%%%%%%%%%%%%%%%%%%%%%%%%%%%%%%%%%%%%%%%%%%%%%%%%%%%5

\subsection{Summary of main results}

Before delving into the technical details of the analysis, it is worth listing the main results that we find in this paper.  The model of three coupled helical edges is 
developed in section \ref{sec:model}, and illustrated schematically in Fig.~\ref{fig:energies_simp}.  The non-interacting system consists of three helical 
edge modes, which could arise by stacking quantum-spin-Hall insulators 
(see \cite{Chalker1995, Balents1996, Kim2016} for related discussions in the context of quantum Hall systems), or alternatively from the reconstruction of 
edge states in a single quantum spin-Hall insulator (which is known to occur also in quantum Hall systems, see e.g. \cite{Chklovskii1992,Wang2017}). 
The essential feature is that in the clean non-interacting system, there are three helical modes, from which one of them is topologically protected against localisation due to 
the intrinsic $\mathbb{Z}_2$ topology dictated by TR symmetry of the model.

Our results consider the fate of this system when weak interactions are introduced. We find that two distinct phases may develop, corresponding to:
\begin{enumerate}
\item An emergent topological (ET) state, whose topology differs from the intrinsic topology of three channels.
In this ET state all three edge modes are protected against localisation when disorder is added to the system, meaning that the low temperature conductance is $G=3e^2/h$; and
\item A state that is characterised by time reversal symmetry breaking (TRSB) in the ground state which destroys the intrinsic topology (that was protected by TR symmetry) 
and leads to a vanishing low-temperature conductance.
\end{enumerate}
The different phases of the system are determined by the relative strength of the intra- and inter-mode interactions. The phase diagram of the model is displayed in 
Figs. \ref{fig:phase_diagram} and \ref{fig:more_gen_PD} later in the paper, and shows that the generic scenario of intra-chain interactions being repulsive and stronger than inter-chain 
interactions (which are also repulsive) corresponds to the TRSB phase.  However, the phase diagram also shows that even within purely repulsive interactions, either 
phase is possible in the presence of tunneling between the channels, indicating that details of the edge in any given realisation of the system are crucial
to determine the fate of the interacting system.

The TRSB and the ET phases share some commonalities. Their low energy excitations (in the clean system) correspond to a gapless plasmon mode, and neutral 
excitations with a gap $\Delta_n$. Both states display the phenomenon of dynamical symmetry enhancement, whereby the symmetry of the ground state is higher than in the 
original problem. Both fixed points can be obtained via an adiabatic deformation of the SU(3) Gross-Neveu model, which ultimately has a $\mathbb{Z}_3$ symmetry.   
It is worth stressing that this is true, even through the microscopic model {\bf does not} possess this symmetry.

We now summarise the physical properties of each of the states.  Firstly, in the TRSB state:
\begin{enumerate}
\item The ground state can be described by quasi long range order parameters. The dominant one is controlled by details of the interaction and can be either 
two-particle or trionic.  One can picture this state as a sliding charge-density-wave.
\item Non-magnetic impurities in the TRSB phase become spontaneously magnetic, due to the spontaneous breaking of TR symmetry in the groundstate.
This means that an initial impurity that is even under TR symmetry, acquires an odd part when the system enters the TRSB phase.
This mechanism and the presence of impurities renders the phase insulating at low energies. This is schematically shown in Fig. \ref{fig:conductance_topological}.
\item In the clean TRSB phase, the system possess a gapless plasmon mode that renders all order parameters quasi-long-range-ordered. 
In the presence of disorder or an appropriate Umklapp scattering if the Fermi-momenta of the different modes have the correct commensurability relationship,
the charge mode is gapped and the order parameter becomes non-zero.
\end{enumerate}
Turning now to the phase with emergent topology
\begin{enumerate}
\item The ground state is a $\mathbb{Z}_3$ symmetry protected topological state, where we stress again that the $\mathbb{Z}_3$ symmetry is itself emergent and therefore 
the lattice model itself is not required to (and in general does not) have this symmetry.
\item The phase boundary between the ET phase and the TRSB phase is described by a critical theory that belongs to the same 
universality class as the three-state Potts model, corresponding in the continuous limit to a conformal field theory (CFT) with central charge $c=4/5$ and 
$\mathbb{Z}_{3}$ parafermionic low energy modes.
\item At temperatures above the neutral gap, the conductance may drop below $3e^2/h$, while it will recover to the full quantum conductance $G=3e^2/h$ at low 
temperature. A schematic diagram of this is plotted in Fig. \ref{fig:conductance_topological}.
\end{enumerate}

All the previous points highlight that while the characterisation of the conductance in the ground state of each phase is an obviously important property to analyse, 
it does not capture all the physical features of the system.

%%%%%%%%%%%%%%%%%%%%%%%%%%%%%%%%%%%%%%%%%%%%%%%%%%%%%%%%%%%%%%%%%%%%%%%%%%%%%%%%%

This article develops as follows: In section \ref{sec:model} we introduce a simple phenomenological model for three helical states in the clean case
that displays the general features, first describing the single particle Hamiltonian, and then introducing generic interactions. In section \ref{sec:boso} we analyse the 
low energy -or infrared (IR)- description of the system in terms of Abelian and non-Abelian bosonization. Here we find that the neutral sector is represented by an 
adiabatic deformation of an emergent SU(3) symmetry. We analyse the structure of all two-particle operators that represent backscattering and introduce the relevant order 
parameters in the TRSB and 
ET phases in \ref{sec:charac}. In section \ref{sec:strongTop} we discuss the stability of the phases against general interaction terms. Following this analysis, in 
section \ref{sec:transition} we discuss the transition between the TRSB and ET phase. To gain further insight we develop an intuition about the structure of the massive 
degrees of freedom in terms of an effective parafermionic model on the lattice that respects all the symmetries of the continuous model. Here we show that in the transition 
region between topological to trivial phase along the edge, a parafermionic mode is trapped in the domain wall. In section \ref{sec:diso} we discuss the fate of disorder in 
the system, showing the difference between these phases. Finally, in section \ref{sec:Discussion} we discuss the results and present our conclusions.

\section{The model}\label{sec:model}
\subsection{Single particle Hamiltonian}

\begin{figure}[ht!]
\includegraphics[width=\linewidth]{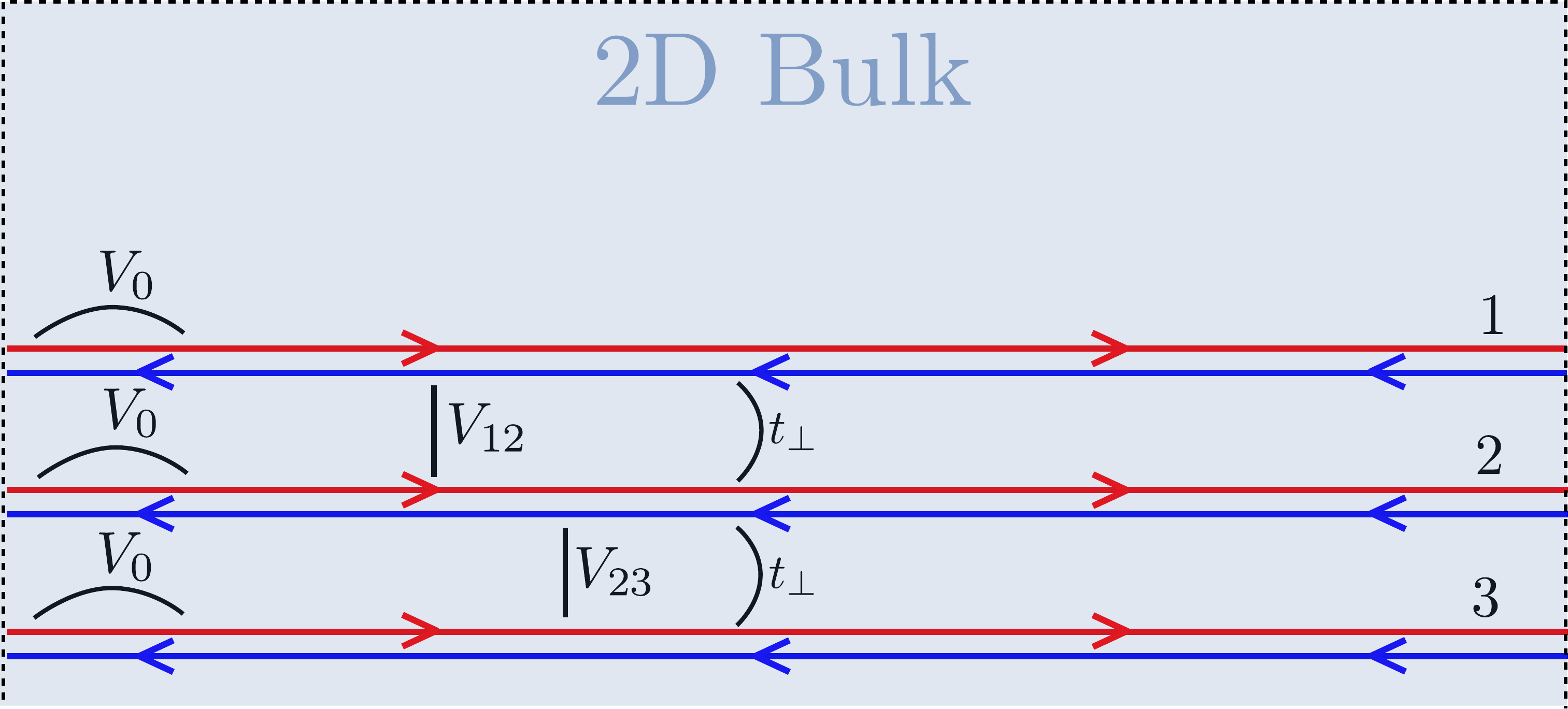} \caption{Color online. Three helical modes on the edge of a two dimensional TI. We label the
different channels by $1,2$, and $3$ and the different interaction strengths $V_0,V_{12},V_{23}$
as depicted. Tunneling amplitude between mode $1-2$  and $2-3$ is denoted by $t_{\perp}$. Tunneling
between 1 and 3 is assumed to be negligible.}
\label{Diagram_simple} 
\end{figure}

While no symmetry apart from TR symmetry should be expected on the edge of a multichannel TR topological insulator, 
to keep the exposition and the relation to the physical regimes clear here we consider a simple model that
displays all the features of the generic model. We analyse a generic model in Appendix \ref{app:General_model}.
We consider three helical modes, described by the fermion destruction and
creation operators of momentum $k$ around the Fermi momenta, $c_{k,a}^{\eta}$ and $(c_{k,a}^{\eta})^\dagger$,
where $a=(1,2,3)$ denotes the mode and $\eta=(+,-)$ labels its helicity.
For small momenta, the non-interacting single particle Hamiltonian is
\begin{equation}\label{sing_part_ham}
H_{0}=\sum_{k,a,\eta}\epsilon_{\eta}(k) n_{k,a}-\frac{t_\perp}{\sqrt{2}}\sum_{k,\eta}\left[(c^{\eta}_{k,2})^\dagger(c^{\eta}_{k,1}+c^{\eta}_{k,3})+\text{h.c.}\right],
\end{equation}
where $\epsilon_{\eta}(k)=\eta v_{F}k$ is the linearized energy of each helical mode. For simplicity we assume that the Fermi velocities $v_F$ of all the modes coincide. 
The operator $n_{k,a}=(c^{\eta}_{k,a})^\dagger c^{\eta}_{k,a}$ measures the number of modes of helicity $\eta$ and momentum $k$. The parameter $t_\perp$ describes the 
tunneling amplitude 
between different modes of the same helicity. 
Here we assume that tunneling only occurs between the modes which
are closest in space. A diagram of the arrangement of helical modes
and their labellings is given in Fig. \ref{Diagram_simple}.
Note that although tunneling between Kramers pairs is forbidden by TR symmetry, tunneling between modes of the same helicity
is not constrained. Generically this tunneling will exists and will be non-universal. In this section we assume that it takes the simple form given by the second term
in Eq. (\ref{sing_part_ham}). A more general tunneling term does not change the overall picture (See Appendix \ref{app:General_model}). 

\begin{figure}
\includegraphics[scale=0.5]{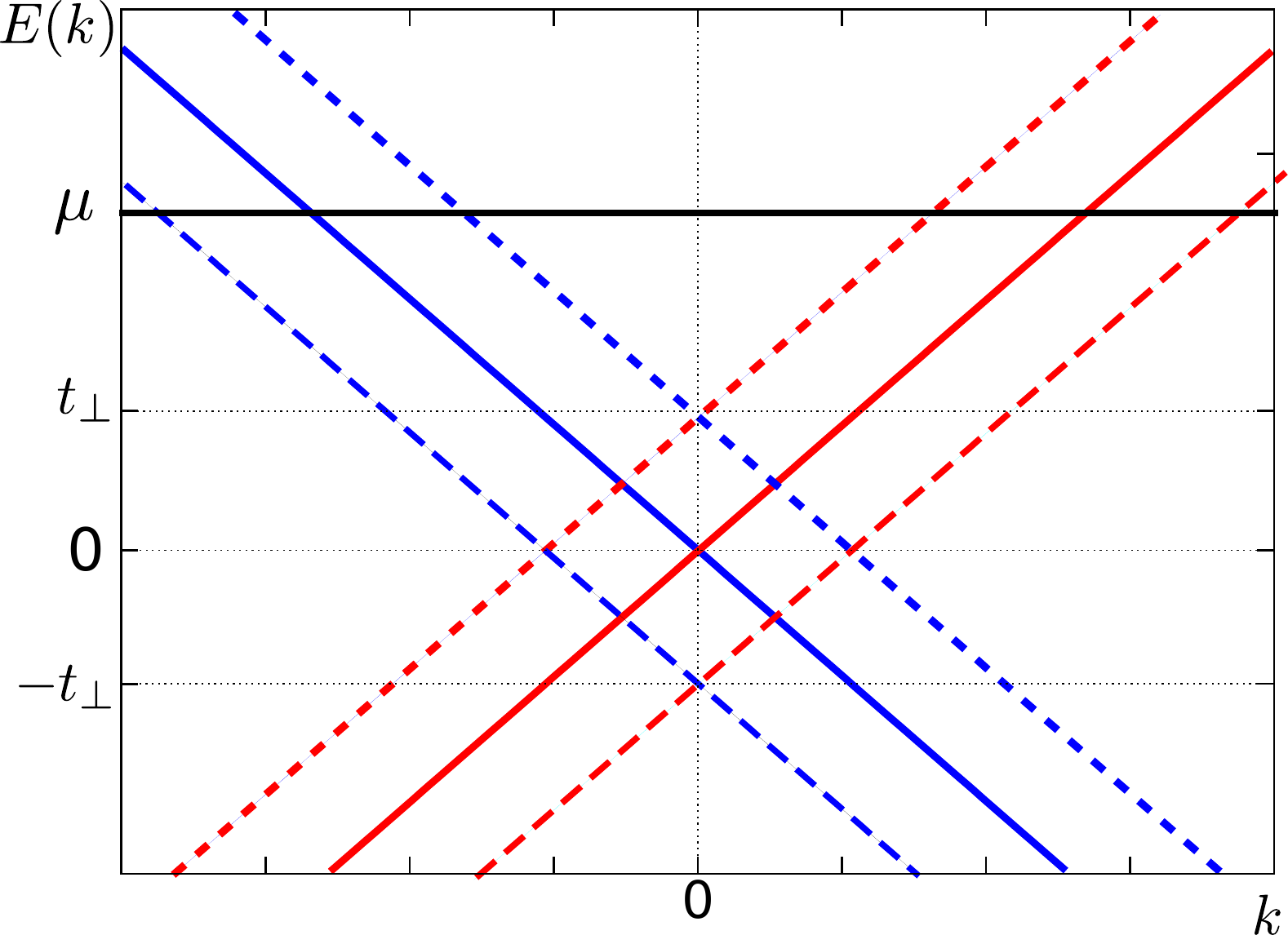} \caption{Color online. Single particle spectrum for the different modes in
the band basis. We consider a linear approximation to the energy spectrum.
Solid lines denote the two chiralities of the mode 2 in the band basis,
while the (short) dashed lines denote the two chiralities of the mode
(1) 3 in the band basis. Note that the crossing points around $k\sim 0$ are protected by TR symmetry, so a gap does not open. All the other crossings are not
protected and in principle energy gaps can be opened. We assume that the chemical potential $\mu$ is such that the Fermi energy does not intersect any crossing point
(represented here by the black horizontal line).
This implies that the low energy physics is captured by three helical modes, where our discussion of the main text applies.}
\label{fig:energies_simp} 
\end{figure}

In the band basis, that corresponds to
\begin{eqnarray}
 \psi_{k,1(3)}^\eta=\frac{c_{k,1}^\eta\pm\sqrt{2}c_{k,2}^\eta+c_{k,3}^\eta}{2},\quad \psi_{k,2}^\eta=\frac{c_{k,1}^\eta-c_{k,3}^\eta}{\sqrt{2}},
\end{eqnarray}
the single particle Hamiltonian is diagonal and the energy dispersion relations are given by
\begin{equation}
E_{\eta}^{a}=\eta v_{F}k+\lambda_{a}t_{\perp},
\end{equation}
with $\lambda_{a}=(-1,0,1)$. These energy dispersions relations are depicted
in Fig. \ref{fig:energies_simp}.

Note that the single particle Hamiltonian is invariant under the symmetry of
interchanging the modes $1\leftrightarrow3$. This symmetry is not expected to hold in general, and we break it explicitly
in the general model of Appendix \ref{app:General_model}.

\subsection{Interactions}

A generic interaction between the three different helical modes is
described by the following lattice Hamiltonian 
\begin{equation}
H_{{\rm int}}=\sum_{i,a}V_{0}n_{i,a}n_{i+1,a}+2\sum_{i}(V_{12}n_{i,1}n_{i,2}+V_{23}n_{i,2}n_{i,3}),
\end{equation}
where the density at each site $i$ and channel $a$ is $n_{i,a}=\sum_{\sigma}(c_{i,a}^\sigma)^{\dagger}c_{i,a}^{\sigma}.$ 
The interaction parameter $V_0$ denotes the intra-mode interaction, while $V_{ab}$ denotes the interaction between modes $a$ and $b$.
For simplicity of the exposition, here we do not consider the interaction between modes 1 and 3, although such interaction is considered in Appendix \ref{app:General_model}.
In the basis that diagonalizes the Hamiltonian,
the density for the band $a$ and helicity $\eta$ corresponds to
$\rho_{i,a}^{\eta}=(\psi_{i,a}^\eta)^{\dagger}\psi_{i,a}^{\eta}$. Summing
over the helicities we have the total density per band $\rho_{i,a}=(\psi_{i,a}^+)^{\dagger}\psi_{i,a}^{+}+(\psi_{i,a}^-)^{\dagger}\psi_{i,a}^{-}$.
The total density on each site is $\rho_{i}=\sum_{a}\rho_{i,a}.$

In the low energy, long wavelength limit, we can introduce a continuous
description of the modes and expand the fields around the Fermi points (here $x=ia_0$, with $a_0$ the lattice spacing)
\begin{equation}
\label{eq_1}
\frac{\psi_{i,a}^\eta}{\sqrt{a_0}}=\psi_{a,\eta}(x),
\end{equation}
together with the slowly varying fields fields $\psi_{a,+}(x)=R_{a}(x)e^{ik_{F}^{a}x}$ and $\psi_{a,-}(x)=L_{a}(x)e^{-ik_{F}^{a}x}$.
Fixing the chemical potential away from the band crossings, and considering $t_{\perp}\neq0$,
the Fermi momenta become $k_{F}^{a}=\frac{\mu+\lambda_at_{\perp}}{v_F}$.
In the continuous description, the non-interacting part of the Hamiltonian is given by 
\begin{equation}
H_{0}=i{v}_{F}\sum_{a}\int dx(R_{a}^{\dagger}\partial_{x}R_{a}-L_{a}^{\dagger}\partial_{x}L_{a}).
\end{equation}
Collecting processes that conserve momentum, (do not have oscillations
with $k_{F}$), the interaction sector of the Hamiltonian becomes
(omitting the space dependence of the densities) $H_{\rm int}=H_{\rho\rho}+H_{\rm nl}$, with
\begin{eqnarray}\nonumber\label{forward}
H_{\rho\rho}&= & \int dx\left(\tilde{V}_{0}\rho^{2}+g\left(\rho_{1}+\rho_{3}\right)^{2}+g'\rho_{2}^{2}+\tilde{g}\sum_{\eta}\rho_{1}^{\eta}\rho_{3}^{\eta}\right.\\%)\\
&+&\left.4g\sum_\eta\rho_{2}^{\eta}(\rho_{1}^{\eta}+\rho_{3}^{\eta})\right),
\end{eqnarray}
containing the forward scattering interaction terms, and
\begin{align}\nonumber
H_{{\rm nl}}= & \tilde{g}\int dx\left(R_{1}^{\dagger}R_{3}L_{1}^{\dagger}L_{3}+L_{3}^{\dagger}L_{1}R_{3}^{\dagger}R_{1}\right)\\\nonumber
+ & 4g\int dx\left(L_{1}^{\dagger}L_{2}R_{2}^{\dagger}R_{3}+L_{2}^{\dagger}L_{3}R_{1}^{\dagger}R_{2}+\mbox{h.c.}\right)\\
+ & 4g\int dx\left(L_{2}^{\dagger}L_{3}R_{2}^{\dagger}R_{3}+L_{1}^{\dagger}L_{2}R_{1}^{\dagger}R_{2}+\mbox{h.c.}\right)\label{NLops}
\end{align}
containing the extra interaction terms. Here $\tilde{V}_0=\frac{a_0}{4}(V_0+V_{12}+V_{23})$ and 
\begin{eqnarray}
g=  \frac{V_0a_0}{8},\,\, g' =  \frac{(V_0-V_{23}-V_{12})a_0}{4},\,\, \tilde{g} = 2(g+g'),\label{parameters}
\end{eqnarray}
Note that the full Hamiltonian is invariant under the operation of permuting the modes $1\leftrightarrow3,$ and the interaction
strengths $V_{12}\leftrightarrow V_{23}.$

Taking $g=0$ (or $g_2=0$ in the general model of Appendix \ref{app:General_model}), 
the three helical model reduces to the two helical system studied in Ref. \onlinecite{Santos2016}, plus a forward scattering
interaction with the antisymmetric band mode $\psi_2^\eta$.

In the limit of zero tunneling $t_\perp=0$, there is another operator that conserves momentum, given by 
\begin{equation}
 \mathcal{O}_{t_\perp=0}=\tilde{g}\int dx R_1^\dagger L_1 L_3^\dagger R_3.
\end{equation}
The presence of this operator, together with the other operator involving just the modes 1 and 3 (first line of Eq. (\ref{NLops})), modifies the low energy behaviour of the 
model, preventing the opening of a gap between the modes 1 and 3, as can be observed in the case of two helical modes \cite{Santos2016},
This result is in line with the intuition that independent helical modes interacting through their densities, away from commensurate filling, are not gapped by interactions.

\section{Bosonization analysis}\label{sec:boso}

We represent the slow part of the fermionic operators as vertex operators
of a bosonic field, as is standard in bosonization \cite{GiamarchiBook2003,GogolinBook2004},
by $R_{a}(x)=\frac{\kappa_{a}}{\sqrt{2\pi a_{0}}}e^{i\sqrt{4\pi}\phi_{R,a}(x)}$,
and $L_{a}(x)=\frac{\kappa_{a}}{\sqrt{2\pi a_{0}}}e^{-i\sqrt{4\pi}\phi_{L,a}(x)}$. Here
$\kappa_{a}$ is a Klein factor satisfying $\{\kappa_{a},\kappa_{b}\}=2\delta_{ab}$. The bosonic
fields satisfy the equal time commutation relations $[\phi_{\eta,a}(x),\phi_{\eta',b}(y)]=\frac{i}{4}\eta\delta_{ab}\delta_{\eta\eta'}{\rm sign}(x-y)$,
with $\eta=(+,-)=(R,L).$
Using these conventions the bosonized form of the density in band $a$ and with helicity $\eta$ is $\rho_{a}^{\eta}=\frac{1}{\sqrt{2\pi}}\partial_{x}\phi_{\eta,a}.$

It is useful to define the following fields 
\begin{align}\label{field_def}
\tilde{\phi}_{\eta,c}(x)=\sum_{a=1}^{3}\frac{\phi_{\eta,a}(x)}{\sqrt{3}},\quad\tilde{\phi}_{\eta,\mu}(x)=\sum_{a=1}^{3}d_{a}^{\mu}\phi_{\eta,a}(x),
\end{align}
together with the inverse relation $\phi_{\eta,a}(x)=\frac{1}{\sqrt{3}}\tilde{\phi}_{c\eta}(x)+\sum_{\mu}d_{a}^{\mu}\tilde{\phi}_{\eta,\mu}(x)$.
The vectors $\bm{d}$ correspond to the three vertices of an
equilateral triangle, see Fig. \ref{fig:arrows}, and are explicitly given by
\begin{equation}
\bm{d}_{1}=\begin{pmatrix}\begin{array}{c}
\frac{1}{\sqrt{2}}\\
\frac{1}{\sqrt{6}}
\end{array}\end{pmatrix},\,\,\bm{d}_{2}=\begin{pmatrix}\begin{array}{c}
0\\
-\frac{2}{\sqrt{6}}
\end{array}\end{pmatrix},\,\,\bm{d}_{3}=\begin{pmatrix}\begin{array}{c}
-\frac{1}{\sqrt{2}}\\
\frac{1}{\sqrt{6}}
\end{array}\end{pmatrix}.
\end{equation}
They satisfy $\bm{d}_{a}\cdot\bm{d}_{b}=\delta_{ab}-\frac{1}{3}.$

\begin{center}
\begin{figure}[t]
\includegraphics[scale=0.4]{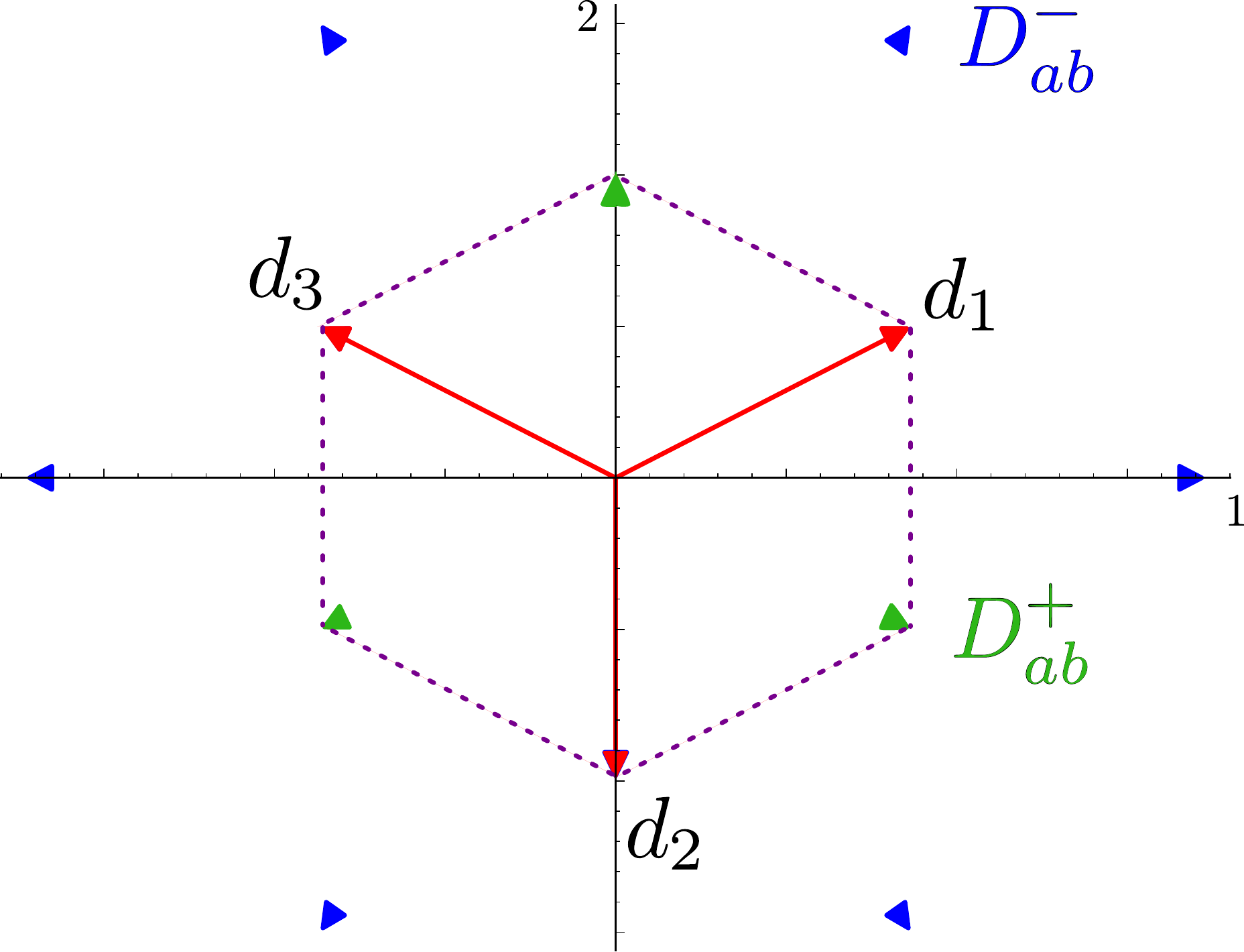} \caption{Color online. Vectors defining the neutral fields and the different order parameters. The vectors of neutral fields form an equilateral triangle (red arrows). 
The horizontal (vertical) axis corresponds to the direction of the neutral field $\tilde{\phi}_{\eta,1} (\tilde{\phi}_{\eta,2})$. These arrows
generate a lattice of possible processes. The lattice points corresponding to $\bm{D}_{ab}^{+(-)}=\bm{d}_a+(-)\bm{d}_b$ are depicted in green (blue). Note that
a process containing any combination of $\bm{D}_{ab}^-\cdot\bm{\tilde{\theta}}$ always contains the field $\tilde{\theta}_1$. This makes the superconducting
order parameters short ranged.}\label{fig:arrows}
\end{figure}
\par\end{center}

We introduce the the non-chiral fields fields $\tilde{\varphi}_{a} = \tilde{\phi}_{R,a}-\tilde{\phi}_{L,a}$ and 
 $\tilde{\theta}_{a}=\tilde{\phi}_{R,a}+\tilde{\phi}_{L,a},$ 
with {$a=c,1,2$}. The only non-vanishing commutation relations in this basis are $[\partial_{x}\tilde{\varphi}_{a}(x),\tilde{\theta}_{b}(y)]=i\delta_{ab}\delta(x-y)$.
For future reference we also introduce the basis for neutral fields $\tilde{\bm{\varphi}}=(\tilde{\varphi}_{1},\tilde{\varphi}_{2})$ and 
$\tilde{\bm{\theta}}=(\tilde{\theta}_{1},\tilde{\theta}_{2})$.

In order to identify the total charge mode we perform a global U(1) transformation on the original fermionic fields 
$\psi_{a,\eta}(x)\rightarrow\psi_{a,\eta}(x)e^{i\Theta}$ which amounts to a shift in the  
bosonic fields as $\phi_{\eta,a}\rightarrow\phi_{\eta,a}+\frac{\Theta}{\sqrt{4\pi}}$. The fields defined in \eqref{field_def} transform as
$\tilde{\phi}_{\eta,c}\rightarrow \tilde{\phi}_{\eta,c}+\sqrt{\frac{3}{4\pi}}\Theta$ and $\tilde{\phi}_{\eta,\mu}\rightarrow \tilde{\phi}_{\eta,\mu}$. This implies that
the fields $\tilde{\theta}_{c},\tilde{\varphi}_{c}$ describe the total charge mode and its conjugate field,  while the modes $\tilde{\theta}_{1,2}$ and their conjugates 
are neutral with respect to the total U(1) charge. 

The Hamiltonian of the system $H=H_0+H_{\rho\rho}+H_{\rm nl}$ in the bosonized variables splits into $H=H_c+H_1+H_2+H_{\rm mix}$, where the total charge sector
$H_c$ is 
\begin{equation}
 H_{c}=\frac{v_{c}}{2}\int dx\left(K_{c}(\partial_{x}\tilde{\varphi}_{c})^{2}+\frac{1}{K_{c}}(\partial_{x}\tilde{\theta}_{c})^{2}\right),
\end{equation}
while the Hamiltonians for the neutral sectors $H_1$ and $H_2$ are
\begin{eqnarray}\label{H_1}
H_{1}&=&\frac{v_{1}}{2}\int dx\left((\partial_{x}\tilde{\varphi}_{1})^{2}+(\partial_{x}\tilde{\theta}_{1})^{2}\right)\\\nonumber
&+&\frac{g+g'}{(\pi a_0)^2}\int dx\cos(\sqrt{8\pi}\tilde{\varphi_{1}}),\quad \mbox{and}
\end{eqnarray}
\begin{eqnarray}\label{H_2}
&H_{2}&=\frac{v_{2}}{2}\int dx\left(K_{2}(\partial_{x}\tilde{\varphi}_{2})^{2}+\frac{1}{K_{2}}(\partial_{x}\tilde{\theta}_{2})^{2}\right)\\
& + & \frac{4g}{(\pi a_{0})^{2}}\int dx\left(\cos(\sqrt{6\pi}\tilde{\varphi}_{2})+\cos(\sqrt{6\pi}\tilde{\theta}_{2})\right)\cos(\sqrt{2\pi}\tilde{\varphi}_{1}),\nonumber
\end{eqnarray}
respectively.
The renormalized velocities and Luttinger parameters of these modes satisfy $v_{1} = v_{F}-\frac{g+g'}{2\pi}$,
\begin{eqnarray}
v_{2}K_{2} & = & {v}_1+\frac{2(g'-g)}{3\pi},\quad
v_{2}K_{2}^{-1}  =  {v}_1+\frac{4g'}{3\pi},\\\nonumber
v_{c}K_{c} & = & {v}_{F}+\frac{g'+5g}{3\pi},\quad
v_{c}K_{c}^{-1}  =  {v}_{F}+\frac{2g'}{3\pi}+\frac{3(g+\tilde{V}_{0})}{\pi}.
\end{eqnarray}
Note that the mode $\tilde{\theta}_1$ sees its velocity renormalized, but its Luttinger parameter stays unity, as a consequence
of TR symmetry and the fact that the microscopic degrees of freedom are helical. This implies that at all orders in the interaction parameters 
the scaling dimension $\Delta_{1}^{\varphi}(\alpha)\equiv\Delta[\cos(\sqrt{2\alpha\pi}\tilde{\varphi}_{1})]=\alpha$.
The remaining part of the Hamiltonian is 
\begin{equation}
 H_{{\rm mix}}=\frac{\sqrt{2}}{6\pi}\int dx\left[(g'-g)\partial_{x}\tilde{\varphi}_{c}\partial_{x}\tilde{\varphi}_{2}+(3g-g')\partial_{x}\tilde{\theta}_{c}\partial_{x}\tilde{\theta}_{2}\right],
\end{equation}
It couples the total charge mode and the second neutral sector. This term is strictly marginal and does not influence the physics
in any of the gapped phases, as the field $\tilde{\varphi}_2$ $(\tilde{\theta}_2)$ is locked by the renormalization of the cosine terms in the ET (TRSB) phase. To first order
in the interactions parameters the scaling dimensions of the cosine terms are
\begin{eqnarray}
\Delta_{2}^{\varphi}\equiv\Delta[\cos(\sqrt{6\pi}\tilde{\varphi}_{2})]=\frac{3}{2}+\frac{g+g'}{\pi v_{F}},\\
\Delta_{2}^{\theta}\equiv\Delta[\cos(\sqrt{6\pi}\tilde{\theta}_{2})]=\frac{3}{2}-\frac{g+g'}{\pi v_{F}}.
\end{eqnarray}

The value of the scaling dimensions determines the fate of the cosine operators under renormalization group (RG).
We  now consider two limiting  cases of purely attractive and purely repulsive interaction. We start with the former, assuming $g=g'$ $(V_0=2(V_{12}+V_{23}))$
for simplicity.

\subsection{Attractive Interactions} 

In this case ${g<0}$, and $\Delta_{2}^{\theta}>\Delta_{2}^{\varphi}$, so the cosine operator $\cos(\sqrt{6\pi}\tilde{\varphi}_{2})$ grows
faster than  $\cos(\sqrt{6\pi}\tilde{\theta}_{2})$ under renormalization.
Keeping the maximal set of commuting cosine operators with smallest scaling dimensions, the model becomes a marginal deformation
of the ${\rm SU(3)}$ Gross-Neveu model \cite{Gross1974} and is given by
\begin{eqnarray}
H & = & H_{c}+H_{{\rm SU(3)}}+\frac{\sqrt{2}g}{3\pi }\int dx\partial_{x}\tilde{\theta}_{c}\partial_{x}\tilde{\theta}_{2}\label{Ham_SU3+}\\
 & + & \frac{4g}{3\pi}\int dx(\partial_{x}\tilde{\theta}_{2})^{2}+\frac{2g}{\pi}\sum_{a}\int dx(\partial_{x}\tilde{\varphi}_{a})^{2}.\nonumber 
\end{eqnarray}
Here the SU(3) symmetric sector  is described by
\begin{eqnarray}\nonumber\label{HSU3}
H_{{\rm SU(3)}} & = & \sum_{a=1}^{2}\int dx \frac{v_{1}}{2}\left((\partial_{x}\tilde{\varphi}_{a})^{2}+(\partial_{x}\tilde{\theta}_{a})^{2}\right)\\\nonumber
 & - & \frac{2g}{\pi}\sum_{a}\int dx(\partial_{x}\tilde{\varphi}_{a})^{2}+\frac{2g}{(\pi a)^2}\int dx\cos(\sqrt{8\pi}\tilde{\varphi_{1}})\nonumber \\
 & + & \frac{4g}{(\pi a)^2}\int dx\cos(\sqrt{6\pi}\tilde{\varphi}_{2})\cos(\sqrt{2\pi}\tilde{\varphi}_{1}). 
\end{eqnarray}

As the prefactors of the cosines flow to strong coupling under RG, the energy of this Hamiltonian is minimised for certain constant values of the 
field $\tilde{\varphi}_{1},\tilde{\varphi}_{2}$. This locking  opens a gap in the spectrum of the neutral sector.
In general, the sign of the amplitude in front of the cosine terms determines the structure of the ground state.
In the case that we are considering here, this amplitude is negative, so the fields $(\tilde{\varphi_1},\tilde{\varphi_2})$ lock to the values
$(0, 0)$. As we will show below this phase is topological  due to the pinning of the neutral field $\tilde{\varphi}_2$.
 The topological nature of this phase  is manifested  in  two ways:
(a) in the  stability of a metallic phase against weak  disorder;
(b)  in  domain wall configurations, that host localised fractionalised zero modes.
In this phase TR symmetry is not broken.

Although for the \textquotedblleft simplified" model discussed above, this phase appears just for attractive interactions,  
for a more generic case  (see  Fig. \ref{fig:more_gen_PD}) the topological phase can emerge for purely repulsive interactions
as well.

\subsection{Repulsive Interactions}

In this regime $g>0$ and the scaling dimensions satisfy $\Delta^\theta_2<\Delta^\varphi_2$, making the cosine operator $\cos(\sqrt{6\pi}\tilde{\theta}_{2})$ the most
relevant operator in RG sense. Keeping the largest set of cosine operators that commute with $\tilde\theta_2$, 
the Hamiltonian becomes
\begin{eqnarray}
H & = & H_{c}+\widetilde{H}_{{\rm SU(3)}}+\frac{\sqrt{2}g}{3\pi }\int dx\partial_{x}\tilde{\theta}_{c}\partial_{x}\tilde{\theta}_{2}\label{Ham_SU3-}\\
 & + & \frac{10g}{3\pi}\int dx(\partial_{x}\tilde{\theta}_{2})^{2}+\frac{2g}{\pi}\int dx(\partial_{x}\tilde{\varphi}_{1})^{2}.\nonumber 
\end{eqnarray}
The Hamiltonian $\widetilde{H}_{{\rm SU(3)}}$ can be obtained from (\ref{HSU3}) by the chiral transformation that interchanges 
$\tilde{\varphi}_2\leftrightarrow\tilde{\theta}_2$. %In terms of the band modes, this transformation is explictly
%\begin{equation}\label{chiral_tr}
% \phi_{L,2}\rightarrow \frac{\phi_{L,1}+\phi_{L,3}}{2},\quad \phi_{L,1}+\phi_{L,3}\rightarrow 2\phi_{L,2}.
%\end{equation}

The cosine operator $\cos(\sqrt{6\pi}\tilde{\theta}_{2})$ grows faster
under renormalization opening a gap, locking the value of the field
$\tilde{\theta}_{2}$. The field values $(\tilde{\varphi}_1^*,\tilde{\theta}_2^*)$ that
minimise the energy are given semi-classically by the solutions of the equations
\begin{eqnarray*}
\cos(\sqrt{6\pi}\tilde{\theta}_2^*)+2\cos(\sqrt{2\pi}\tilde{\varphi}_1^*)&=&0\\
\sin(\sqrt{6\pi}\tilde{\theta}_2^*)\cos(\sqrt{2\pi}\tilde{\varphi}_1^*)&=&0,
\end{eqnarray*}
which for a repulsive interaction in the special point $g=g'>0$ are given by $(\sqrt{2\pi}\tilde{\varphi}_1^*,\sqrt{6\pi}\tilde{\theta}_2^*)=(\pm \frac{2\pi}{3},0)$, 
or by $(\sqrt{2\pi}\tilde{\varphi}_1^*,\sqrt{6\pi}\tilde{\theta}_2^*)=(\pm \frac{\pi}{3},\pi)$ with a double degenerate vaccua.
The dominant order parameters in this phase are odd under TR transformations, indicating the onset of a spontaneous breaking of TR in this phase. This phase is not
topologically protected, as disorder or interaction can gap the charge mode.

\subsection{Generic conditions for the appearance of the different massive phases}
\begin{figure}[t]
\includegraphics[scale=0.7]{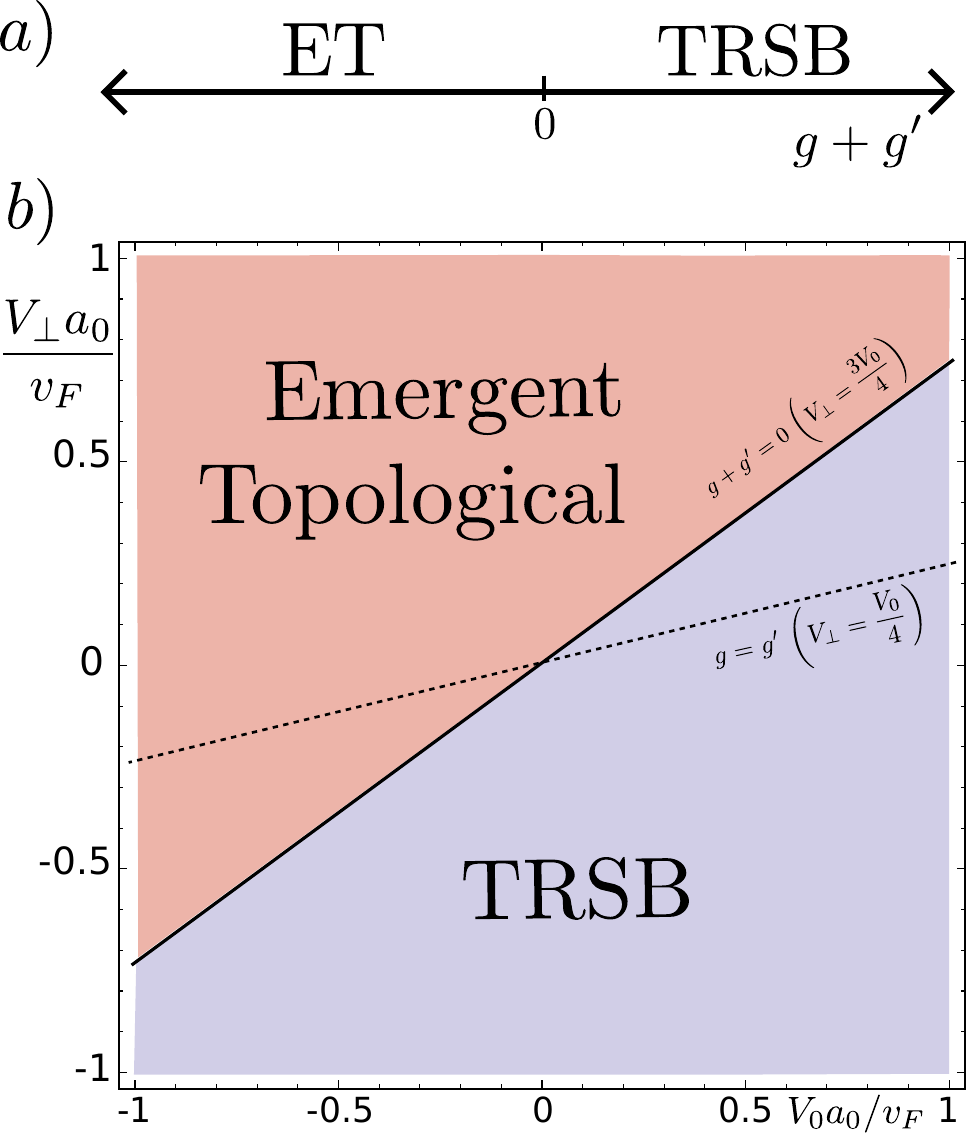} \caption{Color online. a) Different phases as a function of the parameter $g+g'$, defined in 
(\ref{parameters}) (See also Eq. \ref{parameters_fin} in Appendix). b) The phase diagram for the model of section \ref{sec:model}, 
as a function of the intra-mode ($V_{0}$) and inter-mode ($V_{\perp}$) interactions.
Here we assume the same interaction strength between channels  ($V_{12}=V_{23}=V_{\perp}$). 
The transition line between the TRSB and ET phase happens at $g+g'=0$. The diagonal dashed line corresponds to the 
simple limit $g=g'$, considered in Sec. \ref{sec:boso}}
\label{fig:phase_diagram}
\end{figure}

\begin{figure}[t]
\includegraphics[scale=0.5]{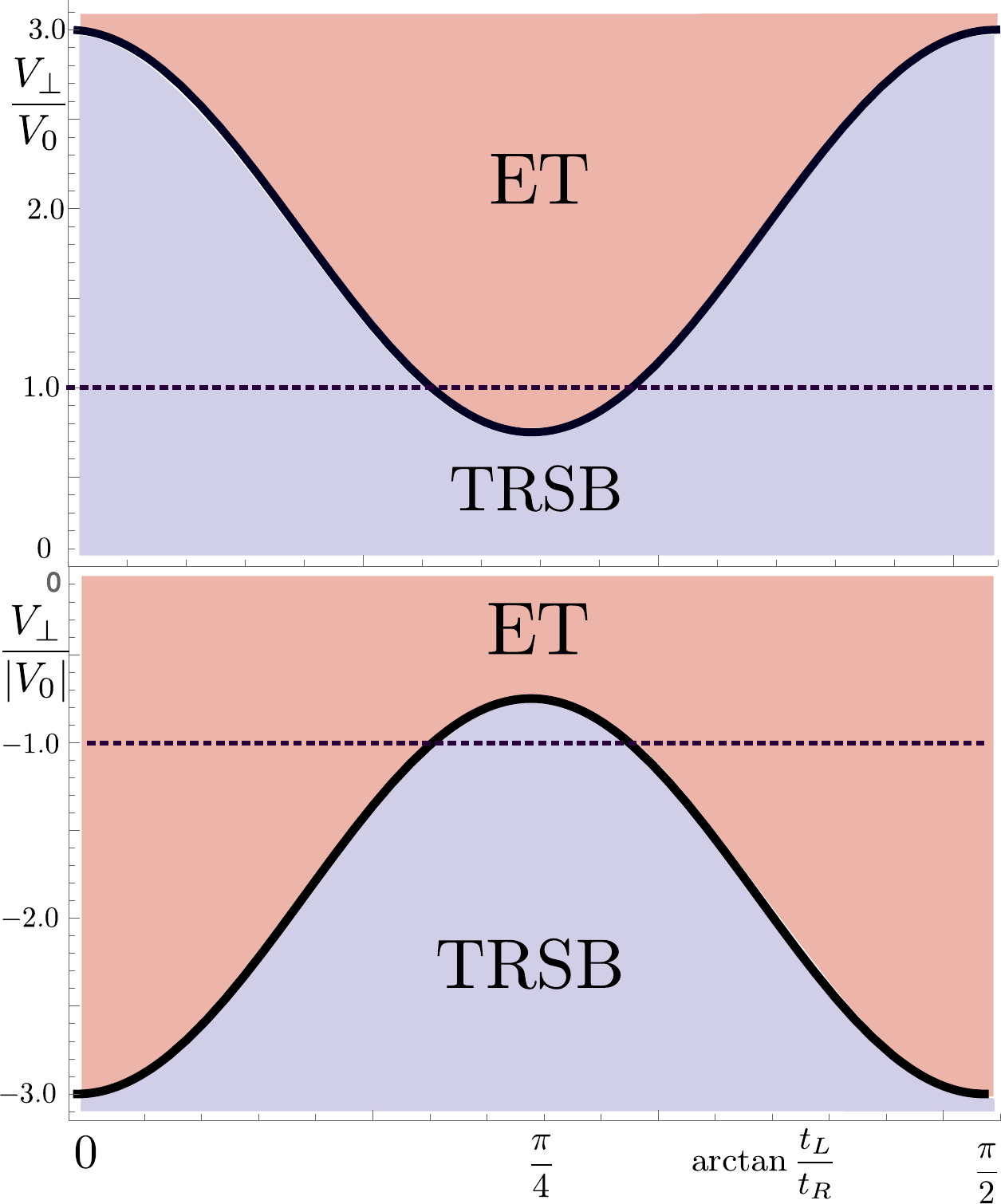} \caption{Color online. In a more general model that includes different tunneling amplitudes between modes 1-2 ($t_L$) and
2-3 ($t_R$) the phase diagram depends on the specific value of the ratio $\frac{t_L}{t_R}$. Assuming the situation $V_{12}=V_{23}=V_\perp<V_0$
the ET phase can be reached for specific repulsive interactions (top panel $V_0>0$) around the symmetric point $t_R\sim t_L$ and generic attractive interactions (lower panel). 
The TRSB phase can be reached with generic repulsive interactions (top panel) and specific attractive interactions (lower panel) around the region of symmetric tunneling.}
\label{fig:more_gen_PD}
\end{figure}

Considering a generic model (see Appendix. \ref{app:General_model}) where we allow for general tunneling amplitudes $t_L$ between modes 1 and 2 and $t_R$ between
modes 2 and 3, we find that both phases can be reached for sufficiently attractive or repulsive interactions, depending on the particular intra- and inter-channel 
interaction strengths. For a simple case of $V_{12}=V_{23}\equiv V_{\perp}$ and $t_R=t_L$, the phase diagram is given by Fig. \ref{fig:phase_diagram}. In the more general
case of arbitrary tunneling amplitudes $t_L$ and $t_R$ and intra-channel interaction larger than inter-channel interaction $V_\perp < V_0$, we find that it is possible to 
reach the ET phase with purely repulsive interactions if the inter-mode tunneling is close to the symmetric case $t_L=t_R$ and the inter-mode interaction $V_\perp$ is 
comparable with the inter-mode interaction strength $V_\perp\sim 3/4 V_0$, see also Fig. \ref{fig:more_gen_PD}.

Below we characterise the ET and TRSB phases in terms of local order parameters.

\section{Characterisation of the Phases}\label{sec:charac}

\subsection{Two Particle Normal Order Parameters}

The usual order parameters involving two-particle number conserving processes
are given by $\mathcal{O}_{\alpha}^{{\rm ord}}=\sum_{ab}(\psi_{a,L}^{\dagger}\lambda_{ab}^{(\alpha)}\psi_{b,R}+\psi_{b,R}^{\dagger}\lambda_{ab}^{(\alpha)*}\psi_{a,L})$,
where $\bm{\lambda}^{(\alpha)}$ are the Gell-Mann matrices \cite{Arfken1995}.
These order parameters can be separated as time reversal even or time
reversal odd by $\mathcal{T}\mathcal{O}_{\alpha,\pm}^{{\rm ord}}\mathcal{T}^{-1}=\pm\mathcal{O}_{\alpha,\pm}^{{\rm ord}}$.
For the even operators we have
that $\lambda_{ab}^{(\alpha)}=-\lambda_{ba}^{(\alpha)}$, while for
the odd operators $\lambda_{ab}^{(\alpha)}=\lambda_{ba}^{(\alpha)}$.
The $3\times3$ antisymmetric hermitian matrices can be generated
by linear combinations of generators of the SU(3) Lie algebra $\bm{\lambda}^{(\alpha)}$ in the fundamental representation (with 
$\alpha\in\alpha_{\rm even}=\{2,5,7\}$)
while the symmetric hermitian $3\times3$ matrices are
generated by linear combinations of $\bm{\lambda}^{(\alpha)}$, with $\alpha\in\alpha_{\rm odd}=\{0,1,3,4,6,8\}$, 
where $\bm{\lambda}^{(0)} $ is the $3\times 3$ identity matrix.

The odd (even) operators under TR are given by $\mathcal{O}_{\alpha}^{{\rm ord}}$, with $\alpha\in\alpha_{\rm odd}$ $(\alpha_{\rm even})$.
The even operators describe the processes of electron hopping that are TR invariant, 
i.e. terms that can be added to the Hamiltonian. The operators that are odd under TR symmetry cannot be included into the Hamiltonian
without explicitly breaking TR symmetry.
Using bosonization, and omitting Fermi momentum contributions, these
operators become, in the basis $\tilde{\bm{\varphi}},\tilde{\bm{\theta}}$
\begin{equation}\label{back_order}
 \mathcal{O}_{\alpha}^{{\rm ord}}=
 \sum_{ab}\tilde{\lambda}_{ab}^{(\alpha)}\cos\left(\sqrt{\pi}\left(\bm{D}_{ab}^{+}\cdot\bm{\tilde{\theta}}+\frac{2\tilde{\theta}_{c}}{\sqrt{3}}\right)\right)e^{i\sqrt{\pi}\bm{D}_{ba}^{-}\cdot\bm{\tilde{\varphi}}},
\end{equation}
where $\bm{D}_{ab}^{\pm}=\bm{d}_{a}\pm\bm{d}_{b}$ (see also Fig. \ref{fig:arrows}). Here we have also incorporated 
the Klein factors $\kappa_{a}$ in the definition $\tilde{\lambda}_{ab}^{(\alpha)}={\lambda}_{ab}^{(\alpha)}\bar{\kappa}	_{a}\kappa_{b}$.
In the TRSB phase (where $\tilde{\theta}_{2}$ is pinned) we observe that the order parameters $\mathcal{O}^{\rm ord}_{4}, \mathcal{O}^{\rm ord}_{5}$ and the combination
$\mathcal{O}_{II}\equiv\frac{i}{2}(\psi_{2,L}^\dagger \psi_{2,R}-\psi_{2,R}^\dagger\psi_{2,L})$ become quasi long-ranged ordered (QLRO).
The correlation function between any of these three order parameters is
$\langle\mathcal{O}_{\alpha}^{{\rm ord}}(x)\mathcal{O}_{\beta}^{{\rm ord}}(0)\rangle\sim{|x|^{\frac{-2K_{c}}{3}}},$
for $\alpha,\beta=(4,5,II)$ and with a wavevector $2k_{F,0}$. The correlation functions between all other two particle normal order parameters decay exponentially.
On the other hand in the ET phase (where the bosonic field $\tilde{\varphi}_{2}$ is pinned) all normal order parameters do not exhibit QLRO and decay exponentially with distance.

\subsection{Superconducting Order Parameters}\label{sec:super}

We can also study the superconducting order parameters,
given by $\mathcal{S}_{\alpha}^{{\rm ord}}=\sum_{ab}(\psi_{a,L}^{\dagger}\lambda_{ab}^{(\alpha)}\psi_{b,R}^{\dagger}+\psi_{b,R}\lambda_{ab}^{(\alpha)*}\psi_{a,L})$.
These operators do not develop QLRO in any phase as they always contain the field
$\tilde{\theta}_{1}$, dual to $\tilde{\varphi}_{1}$, which is locked in both phases (see also Fig. \ref{fig:arrows}). Correlation functions of these order parameters decay exponentially with distance in the 
groundstate. This implies that there is no superconducting order in any of the phases.

\subsection{Trionic order parameters}\label{sec:trionic}

As we have discussed, in both ET and TRSB phases the low energy Hamiltonian of the model corresponds to an adiabatic deformation
of an ${\rm SU(3)}$ Gross-Neveu model. Based on this structure, we can use the fundamental representation of ${\rm SU(3)}$ in terms
of fermions to construct an order parameter. Starting from the complete antisymmetric Young tableaux corresponding to \tiny 
$\Yvcentermath1\yng(1,1,1)$ \normalsize $=c_{1}^{\dagger}c_{2}^{\dagger}c_{3}^{\dagger}$, we define (in the band basis) the order parameter
\begin{eqnarray}
\mathcal{T}_{I} & = & \psi_{1}^{\dagger}\psi_{2}^{\dagger}\psi_{3}^{\dagger}+\text{h.c}.
\end{eqnarray}
with $\psi_a=\psi_{a}^{+}+\psi_{a}^{-}$.
In the TRSB phase the order parameter $\mathcal{T}_I$ acquires QLRO, with correlation function satisfying 
\begin{equation}
\langle\mathcal{T}_{I}(x)\mathcal{T}_{I}(0)\rangle\sim\frac{\sin(k_{3}x)}{|x|^{\frac{1}{2}(\frac{3}{K_{c}}+\frac{K_{c}}{3})}},
\end{equation}
with wavevector $k_{3}=k_{F,3}-k_{F,1}+k_{F,2}$.
This trionic order parameter is dominant for strong attractive interactions such that $K_c>\sqrt{3}\sim 1.7$. We recall that 
for the special point $g=g'$ in the model (\ref{forward}-\ref{NLops}), the trionic order parameter is never more dominant that the two particle
operator $\mathcal{O}^{\rm ord}_{4,5,II}$ of Eq.  (\ref{back_order}). In the general model of Appendix \ref{app:General_model}, we see that 
there is a region where the trionic order parameter is dominant for strong enough interaction.

In contrast, in the ET phase the conjugate field $\tilde{\varphi}_{2}$ is locked. 
This implies that all two-particle order parameters have exponentially decaying expectation values.
In particular, this indicates that the backscattering processes generated by the existence of impurities do not affect the 
conduction properties in this phase, at least at leading order on the impurity strength. 

As there are no two-particle order parameters that dominate in the ET phase, we look for three-particle order parameters.
We find that the operator
\begin{eqnarray}
\mathcal{T}_{II}=\frac{\psi_{1}\psi_{2}^{\dagger}\psi_{3}^{\dagger}+\psi_{1}^{\dagger}\psi_{2}\psi_{3}^{\dagger}+
\psi_{1}^{\dagger}\psi_{2}^{\dagger}\psi_{3}}{\sqrt{3}}+\text{h.c.},
\end{eqnarray}
has dominant correlation function (discarding the purely right/left contributions $R_1^\dagger R_2R_3$, etc.) 
\begin{equation}
\langle\mathcal{T}_{II}(x)\mathcal{T}_{II}(0)\rangle\sim\frac{\sin(3k_{2}x)}{|x|^{\frac{1}{2}(3K_{c}+\frac{1}{3K_c})}}.
\end{equation}
This phase is protected against single particle disorder, and its charge mode cannot be gapped by either two or four fermion terms, regardless of their microscopic
origin. This is a feature of the ET phase. In the following section we discuss on general grounds the topological properties of the TRSB and ET phases.

\section{The stability of topological phases against interactions}\label{sec:strongTop}

So far we have analysed the model of three interacting helical modes, having in mind a microscopic realisation. Now we shift the point of view to a more general
perspective. Here we ask: Once the TRSB or ET phases are fully developed, {\it Is it possible to gap their charge mode, without 
explicitly breaking TR symmetry?.} We ask this question irrespective of any microscopic realisation. For any given model, some of the terms discussed below will
not appear due to momentum conservation or incommesurability. Anyway, they are allowed by the TR symmetry and we consider them. 

In the case of two fermion operators, we have already seen that exist terms that can backscatter the helical modes in the TRSB phase and do not decay exponentially. 
These terms are already present in the non-interacting limit and are the responsible for reducing the classification of two dimensional TR invariant systems from 
$\mathbb{Z}$ to $\mathbb{Z}_2$. For temperatures comparable with the largest gap in the neutral sector, we can estimate their effect by using the non-interacting Landauer 
formula \cite{Landauer1957}, replacing the non-interacting parameters with the renormalized ones, given by the flow of the backscattering amplitudes due to the 
interactions. We do this explicitly in section \ref{sec:diso}. Clearly, for lower temperatures, where the gaps in the neutral sector are the largest energy scale, extended 
backscattering terms {\bf can} gap the charge mode in the TRSB phase, so this phase is {\bf not} topologically protected.
On the other hand, in the ET phase, all two-fermion operators decay exponentially, so they cannot localise the charge mode. 

A general operator allowed by TR symmetry in a system of three helical edges corresponds to a polynomial in the operators
\begin{equation}
 \mathcal{O}_{\bm{n},\delta}^{\theta}=\cos(\sqrt{4\pi}(\bm{n}\cdot \bm{\theta})+\delta),\quad\mathcal{O}^{\varphi}_{\bm{n}}=\exp(i\sqrt{4\pi}\bm{n}\cdot \bm{\varphi}).
\end{equation}
with ${\theta}_a ({\varphi}_a)=\phi_{R,a}+(-)\phi_{L,a}$. Here the parameter $\delta$ is an arbitrary real number and the vector
$\bm{n}$ has integer components. Due to TR symmetry, it satisfies $\sum_a{n}_a=0$ mod 2. 
In the basis of charge and neutral modes, these operators become respectively
\begin{eqnarray}
 \mathcal{O}_{\bm{n},\delta}^{\theta}=\cos\left(\sqrt{4\pi}\left(\frac{2p\tilde{\theta}_c}{\sqrt{3}}+\sum_an_a\bm{d}_a\cdot\bm{\tilde{\theta}} \right)+\delta\right),\\
 \mathcal{O}_{\bm{n}}^{\varphi}=\exp\left(i\sqrt{4\pi}\left(\frac{2p\tilde{\varphi}_c}{\sqrt{3}}+\sum_an_a\bm{d}_a \cdot\bm{\tilde{\varphi}} \right)\right),
\end{eqnarray}
where we have used $\sum_a n_a=2p$, $ p\in\mathbb{Z}$. For example, the superconducting operators $\mathcal{S}^{\rm ord}_{(\alpha_{\rm odd})}$ 
defined in sec. \ref{sec:super} can be written in terms of these general operators as 
$\mathcal{S}^{\rm ord}_{(\alpha_{\rm odd})}\propto \sum_{ab}\lambda_{ab}^{(\alpha_{\rm odd})}\mathcal{O}_{\bm{n}_{ba}^-,0}^{\theta}\mathcal{O}_{\bm{n}_{ba}^+}^{\varphi} $
where we have introduced the vectors $\bm{n}_{ab}^\pm$  defined componentwise as $(\bm{n}_{ab}^\pm)_r\equiv \delta_{br}\pm \delta_{ar}$. The other
superconducting operators can be written similarly. As we mentioned earlier, all of these operators have some contribution from
$(\bm{d}_b-\bm{d}_a)\cdot\bm{\tilde{\theta}}$ which always has a component proportional to $\tilde{\theta}_1$ (see Fig. \ref{fig:arrows}) 
rendering the superconducting operators irrelevant in both gapped phases.

In the TRSB phase, where the pair 
$\tilde{\varphi}_1,\tilde{\theta}_2$ is locked, it is easy to find an operator that locks the charge mode $\tilde{\theta}_c$. A solution (of the infinitely many) is given by
$n_1=n_3=1$ and $n_2=0$, which corresponds to the operator 
\begin{equation}
 \mathcal{O}^\theta_{(1,0,1)}=R_1^\dagger L_1 R_3^\dagger L_3+\text{h.c.}
\end{equation}
In the ET phase, on the other hand, the locked fields are $\tilde{\varphi}_1,\tilde{\varphi}_2$. In this phase we can only use the operator 
$\mathcal{O}^\varphi_{\bm{n}}$ to lock the (conjugate) charge field, as this is the only operator that commutes with the operators that open the neutral gaps. The 
operator $\mathcal{O}^\varphi_{\bm{n}}$ does not conserve the overall charge (because to lock $\tilde{\varphi}_c$ it has to have $p\neq 0$). 
We have encountered operators of this kind in the discussion of trionic order parameters $\mathcal{T}_{II}$ in the ET phase.
Although this operator survives in the ET phase, it cannot be used by itself to lock the charge mode, as it is fermionic in nature and cannot appear
as a term in the Hamiltonian. A valid term that can be included in the Hamiltonian and could serve to lock the charge mode in the ET phase is $\mathcal{T}_{II}(x)\mathcal{T}_{II}(x+a)$.
These results can be summarised as:

{\bf Q}: {\it Is it possible to gap the charge mode in a given phase, without 
explicitly breaking TR symmetry?.}

{\bf A}: In the TRSB phase, it is possible, so this phase is not topologically protected in the presence of interactions. In the
ET phase, on the other hand, it is {\bf not} possible to gap the charge mode, without also breaking particle number conservation, so
this phase is protected by TR symmetry and particle number conservation.
This general analysis implies in particular that the different phases of the microscopic model of three coupled helical wires discussed above are stable under any 
perturbation that does not violate TR symmetry.
This suggests that the model at hand is a representative example for many systems with the same topological properties. It is important to note
that the only way of gapping the charge mode in the ET phase is through locking the $\tilde{\varphi}_c$, which breaks spontaneously TR symmetry if the locking
value is different from zero or $\pi$, as this field is odd under TR and compact.

\subsection{Spontaneous breaking of TR in the trivial phase}

As its name indicates, the TRSB phase breaks spontaneously the TR symmetry in the groundstate. One way of seeing this is by considering 
the expectation value of operators that describe backscattering between Kramers pairs. These are given by the TR odd hermitian operators 
\begin{equation}\label{TR_odd_KR}
\mathcal{O}_{{\rm Kr},a}=\frac{1}{2}\sum_{\eta\eta'}\psi_{a\eta}^{\dagger}(x)(\sigma_{y})_{\eta\eta'}\psi_{a\eta'}(x),
\end{equation}
which in the bosonized form become
\begin{equation}
\mathcal{O}_{{\rm Kr},a}\sim\sin\left(2k_{F,a}x+\sqrt{4\pi}\left(\frac{\tilde{\theta}_{c}}{\sqrt{3}}+\bm{d}_{a}\cdot\tilde{\bm{\theta}}\right)\right).
\end{equation}
The order parameter $\mathcal{O}_{{\rm Kr},2}$
acquires a constant contribution when the charge mode is gapped, which is only possible in the phase where $\tilde{\theta}_2$ is locked.
The order parameters $\mathcal{O}_{{\rm Kr},1}$ and  $\mathcal{O}_{{\rm Kr},3}$ decay exponentially in the TRSB phase.
In particular, this occurs for the microscopic model of section \ref{sec:model} at $\mu=0$ (which corresponds to a commesurability condition that allows
single particle Umklapp scattering) where the operator $\mathcal{O}_{(1,0,1)}^\theta$ conserves momentum and locks the charge mode. 
The presence of a constant order parameter that is odd under TR symmetry indicates the spontaneous breaking of TR symmetry in the groundstate.  

We note that due to the coupling to the charge mode this order parameter has QLRO whenever the charge mode is gapless.
We stress that this consideration is based purely on general grounds and not associated with a particular underlying microscopic model. 
The ET phase, on the other hand, does not break spontaneously TR.

 \subsection{Relation with one and two helical modes}
 
 We observe that the topological protection of the non interacting system can be absent once we include interactions. It is illustrative to consider some simple limits where
 the breaking of non-interacting topological protection is clearly appreciated. Taking $g=0$ in our microscopic model of Eqs (\ref{forward},\ref{NLops}), the system describes
 two strongly interacting modes (modes 1 and 3), coupled just through forward scattering with the mode 2. It shouldn't be surprising that the pair of modes (1,3) can be 
 completely gapped out by disorder, as it is not protected even at the single particle level, (we recall nevertheless, that in the presence of interactions this is possible
 just for repulsive interactions). Let's assume that the pair (1,3) is indeed completely gapped out. By turning on a small $g$ term, the remaining mode is coupled
 to the (1,3) pair, which is localised and acts like an electron puddle. The interaction-induced backscattering with the electrons in this effective puddle breaks the 
 topological protection of the single mode 2, as has been shown in Refs. \onlinecite{Vayrynen2013} and \onlinecite{Vayrynen2014}. Our model reproduces this behaviour.
 
 In the next section, we discuss the nature of the critical line separating the two neutral massive phases.
 
\section{Transition between phases}\label{sec:transition}

In the transition between the TRSB and ET phases, the gap in the neutral sector of the system vanishes throughout the whole edge.
This one-dimensional gapless system is described by a theory at low energies with an emergent $\mathbb{Z}_{3}$ symmetry. By going away from the quantum critical
point, a gap in the neutral sector opens. By considering a position dependent interaction that creates the TRSB phase in one sector of the edge, while 
inducing the ET state on the other, we find that a $\mathbb{Z}_3$ parafermion is trapped in the transition region. 
Below we study the quantum critical point that appears in the transition between these two phases along the edge, and how this result implies 
the existence of nontrivial quasiparticles trapped in domain wall configurations. 
 
\subsection{$\mathbb{Z}_{3}$ critical theory at the transition. }

The transition between the TRSB and the ET phase happens at $g+g'=0$.
The amplitude of the cosine terms in the Hamiltonian (\ref{H_1}-\ref{H_2}) vanishes at the transition in the specific line $g=g'$, indicating that 
along this line of parameters the critical point is Gaussian. By exploring a more generic state e.g. by considering $g\neq g'$ (see also Appendix \ref{app:General_model}), 
the amplitude of the cosine terms does remain finite. On the transition line $g+g'=0$, we find that the Luttinger 
parameters satisfy $K_{1}=K_{2}=1$. This implies in particular that the vertex operators made out of fields $\tilde{\theta}_2$ and $\tilde{\varphi}_2$ are both marginally 
relevant and have the same scaling 
under RG. The competition between these conjugate fields induces a nontrivial fixed point that corresponds to a CFT of central charge $c=4/5$.

We introduce the vertex representation
of the currents of ${\rm {SU(2)}_{1}}$ in terms of the field $\tilde{\phi}_{\eta,1}$
\cite{DiFrancesco1997,FradkinBook2013} 
\begin{equation}
J_{\eta}^{3}=\eta\frac{\partial_{x}\tilde{\phi}_{\eta,1}}{\sqrt{2\pi}},\quad J_{\eta}^{\pm}=J_{\eta}^{1}\pm iJ_{\eta}^{2}=\frac{e^{\mp i\sqrt{8\pi}\tilde{\phi}_{\eta,1}}}{\sqrt{2\pi}},
\end{equation}
which satisfy the Kac-Moody algebra \cite{FradkinBook2013} (repeated
indices are summed over) 
\begin{equation}
[J_{\eta}^{a}(x),J_{\eta'}^{b}(y)]=\eta\frac{i}{4\pi}\delta'(x-y)\delta^{ab}\delta_{\eta\eta'}+i\epsilon^{abc}J_{\eta}^{c}\delta(x-y).
\end{equation}
Using this representation, it is possible to understand the sector
of the Hamiltonian related to the field $\tilde{\phi}_{\eta,1}$ as
a critical SU(2)$_{1}$ Wess-Zumino-Novikov-Witten (WZNW) model \cite{Wess1971,Novikov1982,Witten1983},
perturbed by its primary spin field of scaling dimension $\Delta=\frac{1}{2}$ and a current-current
interaction. In particular, defining the primary field of the WZNW
as $\sigma(x)=e^{i\sqrt{2\pi}(\tilde{\phi}_{R,1}-\tilde{\phi}_{L,1})}$
the Hamiltonian $H$ becomes 
\begin{eqnarray}
H & = & H_{{\rm fs}}+\pi\frac{g+g'}{(\pi a_0)^2}\int(J_{R}^{+}J_{L}^{-}+J_{R}^{-}J_{L}^{+})\\
 & + & \frac{4g}{(\pi a_0)^2}\int(\cos(\sqrt{6\pi}\tilde{\varphi}_{2})+\cos(\sqrt{6\pi}\tilde{\theta}_{2}))(\sigma+\sigma^{\dagger})\nonumber,
\end{eqnarray}
where $H_{\rm fs}$ contains all the forward scattering terms of $H$.
The current-current interaction is a marginal perturbation under RG that vanishes at the transition point, while $\sigma(x)$ is relevant. It will open a gap in the ${\rm SU(2)}_{1}$ sector,
leaving behind a critical Hamiltonian for the $\tilde{\phi}_{\eta,2}$
fields, given by $H\rightarrow H_{{\rm IR}}$ with 
\begin{eqnarray}\label{H_IR}\nonumber
H_{{\rm IR}}&=&\frac{v_2}{2}\int dx\left((\partial_x\tilde{\theta}_2)^2+(\partial_x\tilde{\phi}_2)^2\right)\\
&+&\tilde{g}\int dx \left(\cos(\sqrt{6\pi}\tilde{\varphi}_{2})+\cos(\sqrt{6\pi}\tilde{\theta}_{2})\right),
\end{eqnarray}
and $\tilde{g}$ a non-universal parameter, obtained from the flow
of $g\langle\cos(\sqrt{2\pi}\tilde{\varphi}_1)\rangle$ under RG. This theory corresponds
to a self dual sine-Gordon model, which realises an adiabatic deformation
of an $\mathbb{Z}_{4}$ parafermionic model. This model flows under RG without
opening a gap to an IR fixed point given by a $\mathbb{Z}_{3}$
parafermionic theory \cite{Lecheminant2002}.
As we have seen before, away from the transition line one of the fields ($\tilde{\theta}_2$ or $\tilde{\varphi}_2$) is locked and develops an energy gap. 
This implies that by controlling the interactions spatially, it is possible to go across the quantum phase transition between the two different gapped 
sectors, by moving along the edge. By doing so, we find a parafermionic zero mode trapped in the transition region. These zero modes are studied in the next section.

\subsection{Parafermionic zero modes}\label{sec:para}

As we have found, the transition between the TRSB and the ET
phase is described by a critical theory, whose low energy
description is given by a parafermionic CFT of central charge 4/5,
with $\mathbb{Z}_{3}$ symmetry. Changing the effective interactions
between the helical modes along the edge, for example by external gates, it is possible
to generate a domain wall configuration, where on one side the system is in the TRSB phase, while on the other is in the ET phase.
We can use this result to trap parafermionic quasiparticles in the
interface between the two phases, in a mechanism similar to the
Jackiw-Rebbi fractionalisation of the electron \cite{Jackiw1976}.

Another mechanism to reveal the presence of these parafermionic modes is considering very strong impurity somewhere in the ET region. 
Although it will renormalise to zero at $T=0$, there may be an intermediate energy scale below the scale set by the neutral gap $\Delta_n$ where the impurity 
is still strong and in this intermediate regime one can see the parafermionic edge states (c.f. the equivalent case for two edges discussed 
in \cite{Kainaris2018}).

\begin{center}
\begin{figure}[t]
\includegraphics[scale=0.25]{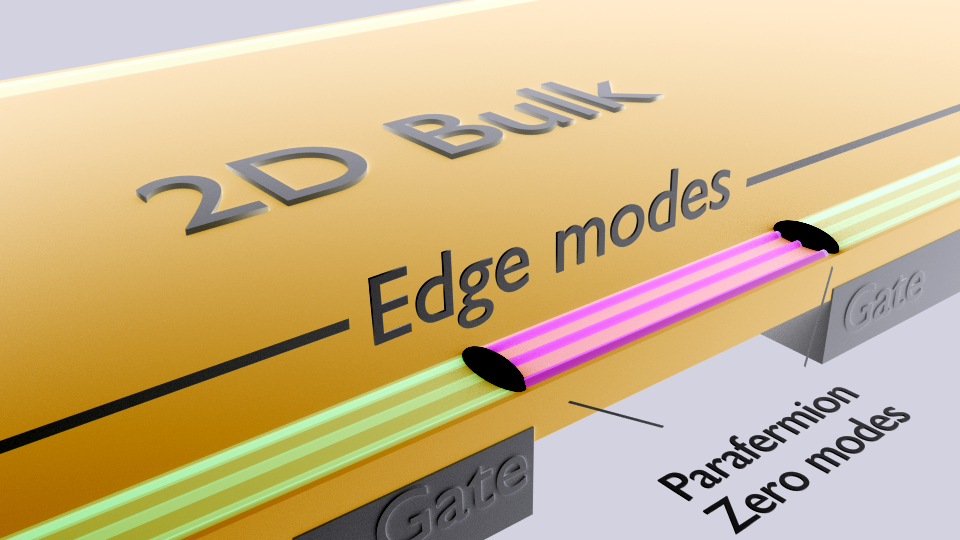} \caption{Color online. By changing the interaction strength along the edge,
the system transitions from the trivial TRSB phase (depicted in yellow) to the ET phase (depicted in pink).
At the boundary between these regions a parafermionic zero mode is localised, represented by the black regions.}
\end{figure}
\par\end{center}

We would like to point out that although the existence of parafermions in one-dimensional gapped fermionic systems has been ruled out in 
\cite{Fidkowski2011}, their existence in quasi one dimensional fermionic gapped systems has been reported in 
\cite{Oreg2014, Klinovaja2014, Klinovaja2014a, Tsvelik2014, Yu2016}. 
In the system considered here, the edge of the two-dimensional TI remains gapless in the ET phase as the system has collective plasmon modes that can be excited with 
arbitrary low energy. This places the edge system discussed in this work in a different category as the quasi-one dimensional systems mentioned previously. Although in 
this system is not possible to gap the charge mode without altering dramatically the ET phase (which is shown to be protected against backscattering), it remains a 
possibility that in similar quasi-one dimensional systems the charge mode could be gapped while maintaining the appearance of parafermionic modes by the emergence of 
$c=4/5$ criticality between two gapped phases. We leave this investigation for the future.

The existence of the gapless charge mode can have an effect on the low energy theory. It could generate hybridization of the edge modes, lifting the zero modes out of zero 
energy by an energy that scales inversely with the system size. Under renormalization this effect corresponds to an irrelevant perturbation that vanishes in the infinite 
size limit. In this sense, the parafermions that we encounter are not protected, although their appear in a topological phase. As their coupling is mediated by the charge 
mode, it would be interesting to look for situations where the charge mode can be completely gapped, while maintaining the structure in the neutral sector that generates 
the parafermions.

To develop some intuition into the nature of these zero modes, we
introduce an effective description on the lattice, following
Ref. \onlinecite{Fendley2012}. 
This lattice description captures qualitatively the physics in the neutral sector, and contains the symmetries
expected to appear around the fixed point obtained from the RG flow of the self-dual Hamiltonian (\ref{H_IR}),
which correspond to $\mathbb{Z}_{3}$ parafermion CFT.

In general $\mathbb{Z}_{n}$ parafermionic modes generalise
Majorana fermions, as they satisfy the relations in the lattice 
\begin{eqnarray}
\chi_{j}^{n}=\eta_{j}^{n}=1,\quad & \chi_{j}^{\dagger}=\chi_{j}^{n-1}, & \quad\eta_{j}^{\dagger}=\eta_{j}^{n-1},\\
 & \chi_{j}\eta_{j}=\omega\eta_{j}\chi_{j},
\end{eqnarray}
where $j$ denotes a lattice site and $\omega=e^{2i\pi/n}$. At different
lattice sites, the parafermions $\eta,\chi$ satisfy 
\begin{equation}
\chi_{j}\chi_{k}=\omega\chi_{k}\chi_{j},\quad\eta_{j}\eta_{k}=\omega\eta_{k}\eta_{j},\quad\chi_{j}\eta_{k}=\omega\eta_{k}\chi_{j},
\end{equation}
for $j<k$. We are interested in a model that captures the symmetry
properties that our system develops in the IR. In particular, the
model should display TR and $\mathbb{Z}_{3}$ symmetry. The simplest
model that displays both is given by the three-state quantum Potts
model, which in terms of parafermions is given by 
\begin{equation}\label{H_eff}
H_{\rm eff}=-\sum_{j=1}^{L}h(\chi_{j}^{\dagger}\eta_{j}\bar{\omega}+\text{h.c.})+J(\eta_{j}^{\dagger}\chi_{j+1}\bar{\omega}+\text{h.c.}),
\end{equation}
with $\bar{\omega}$ the complex conjugate of $\omega$.
The parameters $h,J$ are phenomenological, and represent a description
of the original parameters after renormalization. The phase $\frac{h}{J}\gg1$
corresponds to the ordered phase. In this case the spectrum possess a gap and the
ground state spontaneously breaks the $\mathbb{Z}_{3}$ and TR symmetry.
The opposite limit $\frac{h}{J}\ll1$, corresponds to the disordered
phase, which is also gapped but does not break spontaneously the defining
symmetries. The point $\frac{h}{J}=1$ is critical and self-dual.
The relation with the microscopic parameters for $g>0$ and $g'<0$ is given by 
\begin{equation}
\left[\frac{|g'|}{g}\right]_{{\rm IR}}\sim\frac{J}{h},
\end{equation}
where we denote $[g]_{{\rm IR}}$ the renormalized parameter $g$
in the low energy description. The TRSB phase corresponds to the ordered
phase $\frac{h}{J}\gg1$ (See also discussion at the end of Appendix \ref{app:General_model}). In this phase, the low energy physics is
dominated by the Hamiltonian 
\begin{equation}
H_{\rm triv}=-\sum_{j=1}^{N}\frac{h}{J}(\chi_{j}^{\dagger}\eta_{j}\bar{\omega}+\eta_{j}^{\dagger}\chi_{j}\omega),
\end{equation}
On the other hand, the ET phase corresponds to the limit $\frac{h}{J}\ll1$,
where the Hamiltonian is dominated by 
\begin{equation}
H_{\rm top}=-\frac{J}{h}\sum_{j=1}^{N-1}(\eta_{j}^{\dagger}\chi_{j+1}\bar{\omega}+\chi_{j+1}^{\dagger}\eta_{j}\omega).
\end{equation}
In this phase, the operators $(\Psi_{{\rm in}},\Psi_{{\rm out}})\equiv(\chi_{1},\eta_{N})$
decouple from the Hamiltonian, i.e. $[\Psi_{a},H_{\rm top}]=0$, but they
do not commute with the $\mathbb{Z}_{3}$ symmetry operator $\Omega$,
which has a representation 
\begin{equation}
\Omega=\prod_{j=1}^{N}\eta_{j}^{\dagger}\chi_{j},
\end{equation}
thus satisfying $\Omega\Psi_{a}=\omega\Psi_{a}\Omega$. The zero modes
map states between different symmetry sectors and are localised at
both ends of the topological spatial region.

The TR symmetry $\mathcal{T}$ in this system can be represented as
\begin{equation}
 \mathcal{T}\chi_j \mathcal{T}^{-1}=\eta_{N+1-j},\quad \mathcal{T}\eta_j \mathcal{T}^{-1}=\chi_{N+1-j},
\end{equation}
together with the relation $\mathcal{T}i\mathcal{T}^{-1}=-i$ \cite{Mong2014}.

As we have discussed, a main difference between the ET and the TRSB phase that should be readily accessible in experiments is the value of the conductance.
It is then important to assess the role of disorder in each system. In the next section we analyse the behaviour of a single impurity in each of the phases.

\section{Disorder}\label{sec:diso}
For non-interacting electrons, the conductance through the system is given by Landauer formula
\begin{equation}
G=\frac{e^2}{h}\sum_i T_i,
\end{equation}
where the sum runs over all the transport channels. For the clean system the transmission coefficients $T_i=1$ such that the total conductance through the system 
is $G=3 e^3/h$. In presence of static disorder the problem can be solved using the scattering matrix formalism.
For a single non-magnetic impurity the electric conductance is given by (see also Appendix  \ref{scattering_matrix})
\begin{equation}
\label{conductance}
G(T)=\frac{e^2}{h}\bigg[1+2\frac{1-g_{\rm imp}^2}{1+g_{\rm imp}^2}\bigg].
\end{equation}
The first term on the right-hand side of Eq. (\ref{conductance}) follows from a ballistic propagation along the topologically protected channel.

For an interacting system the Landauer approach is strictly speaking not applicable.
Nevertheless, one may still use it  as a semi-qualitative  approximation. In this case, one needs to replace  the values of transmission coefficients by 
their renormalised value at energy/temperature $T$ (not to be confused with the transmission coefficients $T_i$) dependent scale, $g_{\rm imp} \rightarrow g_{\rm imp}(T)$.
However, Eq. (\ref{conductance}) is valid provided that the system remains in topologically non-trivial state (either inherited or emergent).
If topological protection is removed, the conductance will generically go to zero.

The backscattering processes are in general proportional to the Fermi momentum components of the order parameters $\mathcal{O}_{\alpha}^{{\rm ord}}$ studied previously.
Let's model a single point-like impurity at $x=0$ that backscatters the helical modes by
\begin{equation}\label{impurity}
 \mathcal{O}_{\rm imp}(x)=i\delta(x)(L_1^\dagger R_3-L_3^\dagger R_1) +\text{h.c.},	
\end{equation}
this operator is TR even, so it can be considered as a non-magnetic impurity. The spatially extended version of this operator corresponds to 
$\mathcal{O}^{\rm ord}_{5}$ discussed in Sec. \ref{sec:charac}. In the bosonic language this operator becomes
\begin{eqnarray}\nonumber
 \mathcal{O}_{\rm imp}(x)&=&i\delta(x)\left[\bar{\kappa}_1\kappa_3e^{i\sqrt{2\pi}\tilde{\varphi}_1}-\bar{\kappa}_1\kappa_3e^{-i\sqrt{2\pi}\tilde{\varphi}_1}\right]\\
 &\times& e^{i\sqrt{\frac{2\pi}{3}}[\sqrt{2}\theta_c+\tilde{\theta}_2]}+\text{h.c.}
\end{eqnarray}
where we have written explicitly the Klein factors $\bar{\kappa}_a,\kappa_a$
In the TRSB phase, the fields $(\sqrt{2\pi}\tilde{\varphi}_1^*,\sqrt{6\pi}\tilde{\theta}_2^*)$ are locked into the values $(\pm \frac{2\pi}{3},0)$ or $(\pm\frac{\pi}{3},\pi)$.
The spontaneous choice of any of these configurations in the groundstate breaks TR symmetry as the bosonic field $\tilde{\varphi}_1$ is odd under TR. In this case we observe that a non-magnetic impurity like (\ref{impurity})
becomes $\mathcal{O}_{\rm imp}=\delta(x)(\cos(\sqrt{2\pi}\tilde{\varphi}_1^*)\mathcal{O}_++\sin(\sqrt{2\pi}\tilde{\varphi}_1^*)\mathcal{O}_-)$, where
\begin{equation}
 \mathcal{O}_\pm=ie^{i\frac{\pi}{4}(1\mp 1)}[\bar{\kappa}_1\kappa_3\pm \bar{\kappa}_3\kappa_1]e^{i\sqrt{\frac{2\pi}{3}}[\sqrt{2}\theta_c+\tilde{\theta}_2^*]}+\text{h.c.},
\end{equation}
and  $\mathcal {T}\mathcal{O}_{\pm}\mathcal{T}^{-1}=\pm \mathcal{O}_\pm$. This implies that the spontaneous breaking of TR symmetry in the groundstate creates an effective 
magnetic impurity out of a non-magnetic one. This can be understood as follows: In the TRSB phase the gapless charge mode smears out the TR breaking in the neutral sector, 
such that there is no true long range order parameter and just QLRO. By placing a nonmagnetic impurity, the charge mode is locally pinned to a value that minimizes the 
energy around the impurity. By pinning down the charge around the impurity, the TR breaking of the groundstate is revealed, and the impurity becomes effectively magnetic. 

For temperatures $\Delta_n \ll T \ll \Delta_b$ the charge transport properties of the system are equivalent to the three spinless Luttinger liquids. In this regime a single 
impurity undergoes the standard Kane-Fisher renormalization \cite{Kane1992,Kane1995}
\begin{equation}
\label{rg1}
\frac{d{g}_{{\rm imp}}}{d\ell}=\left(1-\Delta_{\rm imp}\right)g_{{\rm imp}},
\end{equation}
where $\Delta_{\rm imp}=\frac{K_c}{3}+\frac{K_2}{6}+\frac{1}{2}$ is the scaling dimension of the impurity (\ref{impurity}) before the 
neutral gap is opened. Here $\ell=\ln\Delta_b/T$ where we take the bulk gap $\Delta_b$ as the ultraviolet cut-off in this regime.
For an impurity with a weak bare value, the conductance in this range of temperatures will be close but below $G=3e^2/h$,  monotonously decreasing as temperature decreases. 
As the temperature approaches the scale $\Delta_n$ the low energy fixed point where the neutral modes are gapped starts to control the conductance.

In the TRSB phase the impurity operator $\mathcal{O}_{\rm imp}$ survives the integration of the massive
degrees of freedom, and one is left with an effective theory in the gapless charge sector, with the impurity operator now given by
\begin{equation}
 \mathcal{O}_{\rm imp} \propto g_{{\rm imp}} \cos \left( \sqrt{\frac{4\pi}{3}} \theta_c(0) \right).
\end{equation}
For a sufficiently small amplitude $g_{{\rm imp}}$, even after the RG flow discussed before, the renormalized value of the impurity strength
will remain small, such that another Kane-Fisher renormalization analysis can be done. The impurity now scales under RG with the scaling dimension 
$\Delta=K_c/3$ and an ultraviolet cut-off determined by $\Delta_n$.
For any repulsive interaction it is a strongly relevant perturbation. Thus the impurity strength will flow to a strong coupling fixed point, locking the charge field
around the impurity and  
making the conductance vanish at zero temperature. Note that in the TRSB phase the impurity is effectively magnetic, so Eq.~(\ref{conductance}) is no longer valid.

We can say something about this strong coupling limit by constructing the leading irrelevant operator that creates a soliton in the $\theta_c$ field at this point; 
i.e. the operator responsible for non-zero current \cite{Carr2013,Saleur1998}.  This operator is $\cos(\sqrt{12\pi}\varphi_c(0) )$, which has scaling dimension $d=3/K_c$, and hence 
the conductance at low temperature is $G(T) \propto T^{2(d-1)} = T^{6/K_c-2}$.

In contrast,  in the ET phase, after the massive degrees of freedom are integrated out, electron and trion backscattering do not contribute.
Therefore the conductance of the system in the topological phase at low temperatures is almost perfect, $G \approx 3e^2/h$.  While one would usually then analyze 
the approach to perfect conductance in a way similar to above, by finding the leading irrelevant operator present after integrating out the massive degrees of freedom, 
the discussion in Sec. \ref{sec:strongTop} indicates that no such operator exists in the ET phase.  Hence we expect these corrections not to be power law.
The approach to perfect conductance in the ET phase remains an open question.

\begin{figure}[ht!]
\includegraphics[width=\linewidth]{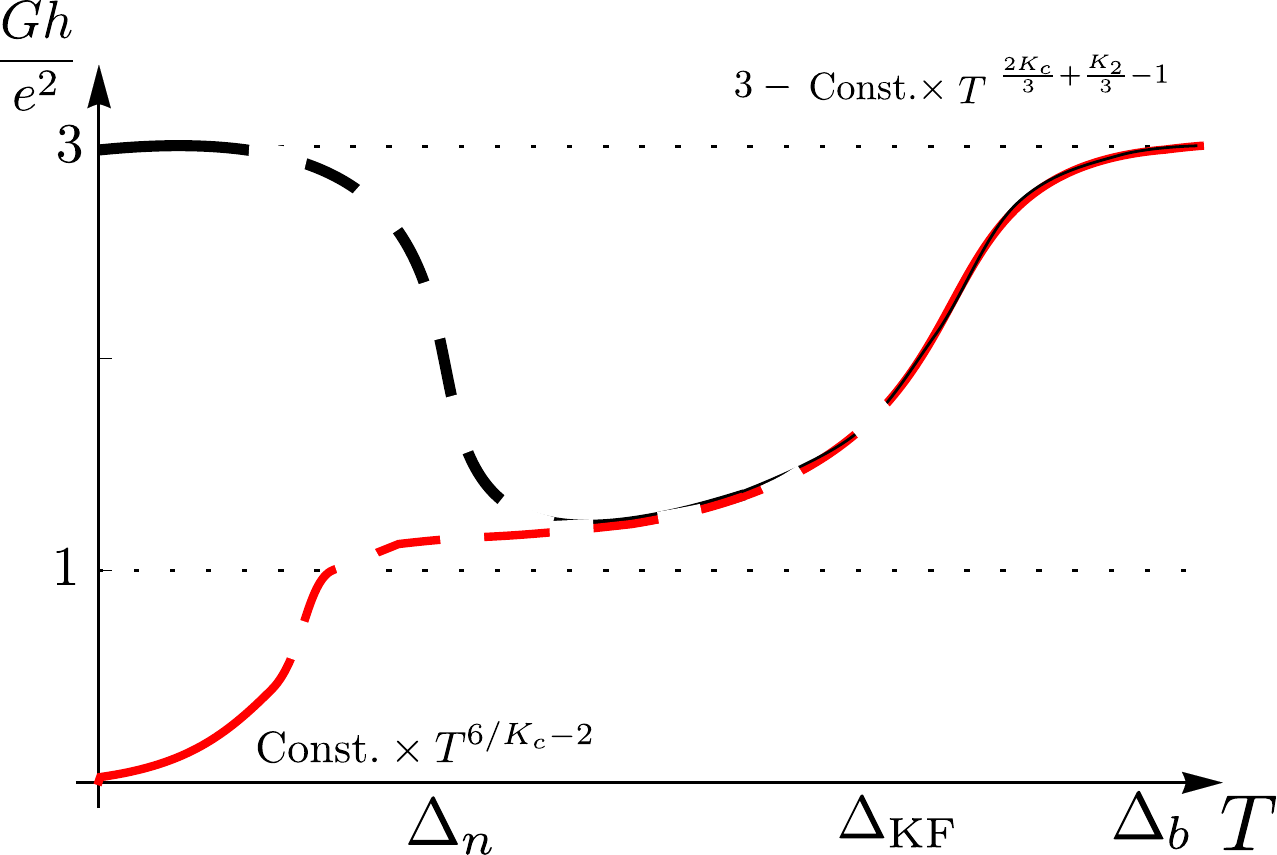} 
\caption{(Color online). Schematic plot of  electric conductance as a function of temperature in the ET phase (black) and in the TRSB phase  (red line).
The different energy scales associated with the gap of the two dimensional TI $\Delta_b$, the gap in the neutral sector on the edge $\Delta_n$
and the scale where the renormalized  value of an impurity becomes of order one $\Delta_{\rm KF}$, determine the behaviour of the conductance, with the corresponding exponents shown.
The transitions between the different regimes are shown in dashed lines, and are just schematic.}
\label{fig:conductance_topological} 
\end{figure}

We now schematically plot the conductance as function of temperature for the both phases, see Fig. \ref{fig:conductance_topological}.
We focus on the limit where the bare  value  of  impurity potential is weak. We assume that the interaction  is repulsive and its strength is small, such that all 
characteristic Luttinger liquid parameters are slightly smaller than one. Furthermore, we assume for definiteness that the energy scale $\Delta_{\rm KF}$ where the impurity becomes strong 
is higher than the energy scale $\Delta_n$. Just below the two dimensional TI's gap $\Delta_b$, the conductance in all phases is a non universal function with a value below (but close to) 3 (in the units of $e^2/h$).
As temperature decreases towards the scale $\Delta_{\rm KF}$ the conductance in both phases decreases as $T^{\frac{2K_c}{3}+\frac{K_2}{3}-1}$.
Between $\Delta_{\rm KF}$ and $\Delta_n$ the system behaves as three helical modes in presence of a non-magnetic impurity. This signals that the
conductance develops a plateau around $G=1$. This plateau extends roughly throughout the range of energies between $\Delta_{\rm KF}$ and $\Delta_n$.
Below $\Delta_n$ the conductance in the ET phase starts to rise with decreasing temperature, reaching an ideal limit $G \rightarrow 3$ at $T\rightarrow0$.
Therefore in this phase the conductance is a non-monotonic function of temperature. 
In the TRSB phase, below the neutral gap the behaviour of the conductance is fully controlled by an insulating fixed 
point, and the conductance approaches zero as $T^{6/K_c-2}$.

In the opposite scenario, when $\Delta_n>\Delta_{\rm KF}$, generically there is no plateau around $G\sim 1$ and the conducting properties of the system are fully 
dictated by the opening of the neutral gap.

\section{Discussion and Outlook}
\label{sec:Discussion}
In this paper we studied the competition of emergent and inherent topological orders. 
We focused on a system made of three helical wires, that may arise as edges states
of three copies of two-dimensional topological insulators stacked together or by the edge reconstruction of a single copy.
In the non-interacting limit this system  is topologically equivalent to a single helical edge state protected against static disorder.
We showed that in the presence of electron interaction this picture changes. 
We now summarise our findings.

In the presence of interaction the system may turn into one of two  possible states.
In the first case the system acquires new topological order that can not be adiabatically connected to the non-interacting one.
In the second case the TR symmetry is spontaneously  broken and the system is driven into a topologically trivial state, 
that becomes an Anderson insulator in the presence of a static disorder.

To understand the loss of topological protection one may take the limit where one of the channels weakly interacting with the rest.
The remaining two channels may be in the topologically trivial or non-trivial state, depending on the interaction strength \cite{Santos2016,Kainaris2015}.
If two coupled channels happen to be in a topologically trivial state, they would be localised by any finite amount of disorder.
Therefore the system of three helical modes effectively becomes equivalent to a single helical channel coupled by hopping to multiple puddles of electronic fluid.
Such system is equivalent to an Anderson insulator \cite{Vayrynen2013,Vayrynen2014}.
 
The ground state of a topologically  trivial state is a strongly correlated one, 
that develops  a QLRO.  The character  of QLRO  depends on the details of interaction.
Weak  repulsive interaction results in a family of  two-particle correlations with power low decay and $2k_F$ oscillations.  
For sufficiently strong repulsive interaction $K_c >\sqrt{3}$ and the dominant  QLRO is a trionic one.
 
In the case of small attractive interactions, a new topological order develops.
The latter is protected by a gap in the neutral sector, that opens inside the one dimensional system due to many body scattering.
This state is robust against Anderson localisation  with a total conductance of $3e^2/h$ for moderate disorder.

The transition between topological and non-topological phases occurs along a line in the parameter space.
While the neutral sector of the theory is gapped at both sides of the line, it becomes gapless at the transition. 
Its low energy behaviour corresponds to the $\mathbb{Z}_3$ parafermionic CFT universality class.
The latter is manifested by the emergence of parafermionic excitations at the end points of the system.

We also find that the low energy fix point has a higher symmetry with respect to interaction between  modes that the 
original model, signalling  a {\it dynamically emergent symmetry}. This phenomenon was previously observed in the context of  three leg ladders
\cite{Lecheminant2008,Rapp2008,Azaria2010,Lecheminant2012,Pohlmann2013,Okanami2014,Sotnikov2015,Nishida2015,Klingschat2010,Miyatake2015,Ozawa2010,Azaria2017}.
In our case, the massive phases  are ground states of a Hamiltonian that is  obtained by   marginal deformations of an emergent SU(3) symmetry, which is not 
present in the UV, but that manifest itself in the IR.  
The topological phase corresponds to a deformed SU(3) Hamiltonian $\tilde{H}_{\rm SU(3)}$ of Eq. 
(\ref{Ham_SU3+}) that can be obtained from the usual SU(3) Gross-Neveu Hamiltonian by performing a chiral transformation.
The emergent topology arises due to a gap in the neutral sector of this Hamiltonian. 

Though both symmetry protected topological ordered and dynamically generated symmetries were previously known, 
the current system is the first example where both effects act together.
The interaction enhances the effective symmetry of the problem in the IR limit. 
The generated symmetry gives rise to the topologically nontrivial state.

The rich physics of this system invites to a further exploration of its different facets.
In particular, we consider crucial to find experimental signatures of parafermions that emerge on the boundary between the phases,
to assess their stability in the presence of a gapless charge mode, and to 
account for strong impurities and random disorder. It is appealing to consider how these results generalise to a larger number of helical modes,
exploring the possible connection to the theory of interacting symplectic wires.
It remains to be seen if the emergent symmetry allows to find the regimes beyond those predicted within disordered Fermi liquid approach \cite{Finkelstein}.
Finally, from a general perspective, it is compelling to study the general criteria for the existence of dynamically emergent symmetry protected states.

{\it Note added:}
When this manuscript was in preparation, we learned
about preprints \cite{Kagalovsky2018,Keselman2018}  with partly overlapping content.  The work of Kagalovsky et al. \cite{Kagalovsky2018} discusses TRS breaking in 
the ground state leading to zero conductance at zero temperature, for any number of channels $N\ge 3$. Our results are in full 
agreement with theirs for $N=3$. In this specialised case we uncover a number of non-trivial phases as function of interaction and crossovers as a function of temperature, 
which presumably one would see for any odd $N$, although to confirm or deny this conjecture remains work for the future.  The work of Keselman et al. \cite{Keselman2018} looks at a different model, 
concentrating on $N=3$ channels in which the non-interacting model is non-topological, and like us finds a phase with TR symmetry breaking, and another phase with an emergent 
topology.  While their TRSB phase is the same one that we find, they curiously find a different emergent topological phase, in the universality class of the Haldane 
spin-1 chain as opposed to our $\mathbb{Z}_{3}$ parafermionic state.  This gapless Haldane state relies on a $(\mathbb{Z}_2)^3$  symmetry, which we explicitly break by the 
inter-chain hopping (or equivalently, the splitting of the Fermi-points) in our model.  In contrast, our parafermion state explicitly emerges from interaction terms that
require the inter-chain tunneling in the Hamiltonian. It remains work for the future to determine the full phase diagram of a more generic $N=3$ channel system, and to see 
if there are more possibilities for emergent topological states beyond these two.

{\it Acknowledgement}.- R.S. would like to thank Eran Sagi, Jinhong Park and Benjamin B\'eri for stimulating discussions.
D. G. was supported by  ISF (grant 584/14) and  Israeli
Ministry of Science, Technology and Space. R.S. acknowledges funding from by EPSRC grant EP/M02444X/1,
and the ERC Starting Grant No. 678795 TopInSy.

\begin{appendix}

\section{General Model}\label{app:General_model}

\begin{figure}[ht!]
\includegraphics[scale=0.4]{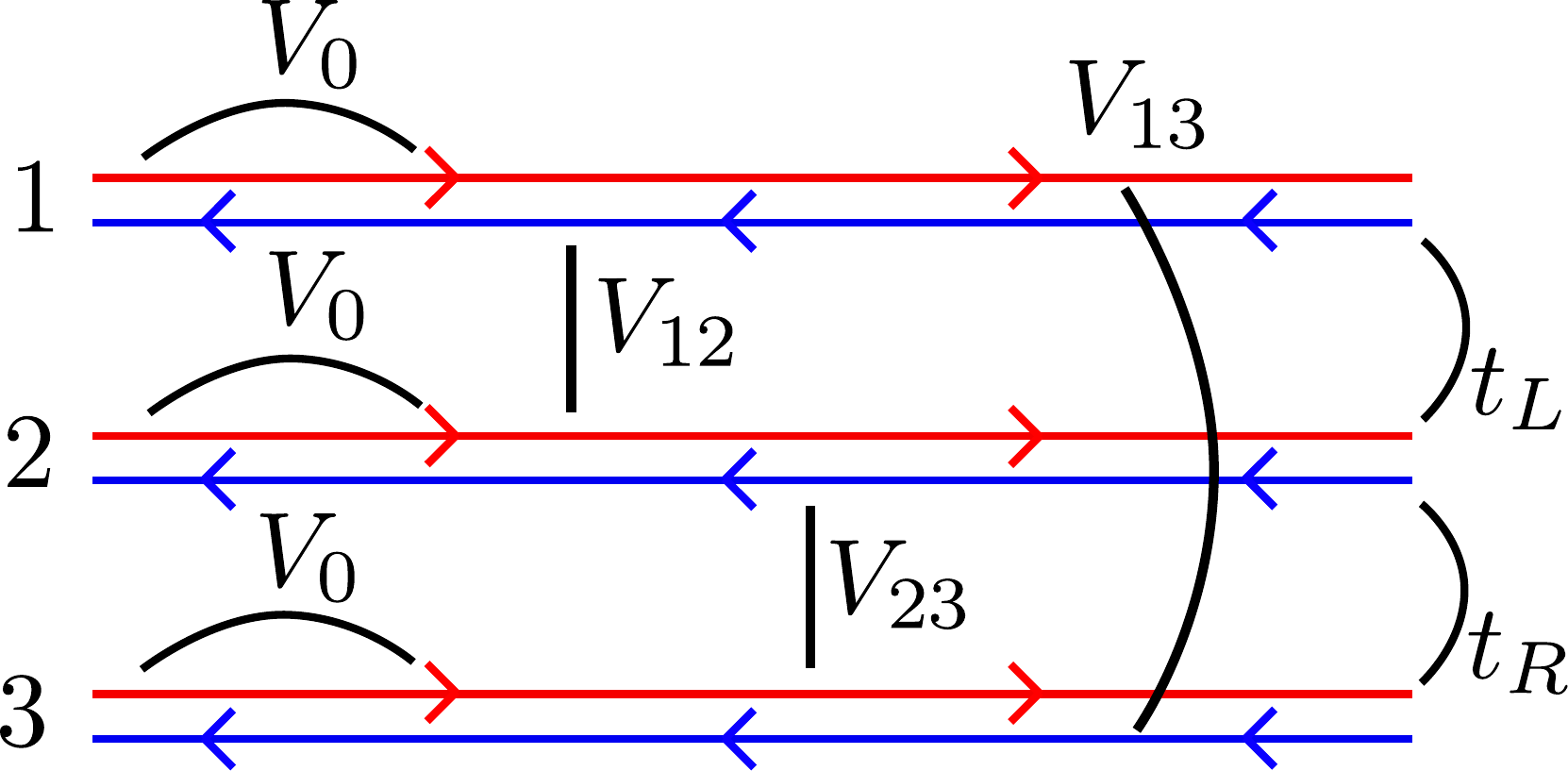} \caption{Color online. Generic diagram of three helical modes. We label the
different channels by 1,2, and 3 and the different interaction strengths
as depicted. Tunneling amplitude between mode 1 and 2 is denoted $t_{L}$,
while tunneling between 2 and three is denoted $t_{R}$. Tunneling
between modes 1 and 3 is assumed to be negligible.}
\label{Diagram} 
\end{figure}

The model considered in the main text corresponds to a particularly simple description of a more generic model that we discuss here.
Using the same notation as the main text, we consider three helical modes, described by the fermion destruction
operator of momentum $k$, $c_{k,a}^{\eta}$,
where $a=(1,2,3)$ denotes the mode and $\eta=(+,-)$ labels its helicity.
For small momenta, the non-interacting Hamiltonian is
\begin{eqnarray}\nonumber
H_{0}&=&\sum_{k,a,\eta}\eta v_{F} k (c^{ \eta}_{k,a})^\dagger c^{\eta}_{k,a}+\alpha_{so}k (c^{\eta}_{k,a})^\dagger c^{\bar{\eta}}_{k,a}\\
&-&\sum_{k,\eta}(t_L (c^{\eta}_{k,2})^\dagger c^{\eta}_{k,1}+t_R(c^{\eta}_{k,2})^\dagger c^{\eta}_{k,3}+\text{h.c.}),
\end{eqnarray}
where $v_{F}$ is the Fermi velocity of the modes, $\alpha_{so}$
parameterizes a residual spin-orbit coupling along the edge. We assume that tunneling only occurs between the modes which
are closest in space, with amplitudes $t_L$ and $t_R$. A diagram of the arrangement of helical modes
and their labellings is given in Fig. \ref{Diagram}. 

The energy dispersion relations in the band basis are $E_{\eta}^{a}=\eta\tilde{v}_{F}k+\lambda_{a}t_{\perp},$
with the new Fermi velocity $\tilde{v}_{F}=\sqrt{v_{F}^{2}+\alpha_{so}^{2}}$,
the perpendicular tunneling parameter $t_{\perp}=\sqrt{t_{L}^{2}+t_{R}^{2}}$
and $\lambda_{a}=(-1,0,1)$. The single particle Hamiltonian is invariant under the symmetry of
interchanging the modes $1\leftrightarrow3$ and $t_{L}\leftrightarrow t_{R}$.

Going from the original modes to the band modes that diagonalize the Hamiltonian is implemented by the unitary
transformation $[U(v)]_{ab}^{\eta\eta'}=(U_{{\rm c}}(v))_{ab}(U_{{\rm h}})^{\eta\eta'}$
where $\tan v=\frac{t_{L}}{t_{R}}$. The unitary transformation $U_{{\rm h}}=e^{i\beta\sigma_{y}}$
(with $\tan2\beta=\frac{\alpha_{so}}{v_{F}}$) acts on the helicities,
while $U_{{\rm c}}$ acts in the channel index rotating the modes
into the band basis, and is given by 
\begin{equation}
U_{{\rm c}}(v)=\frac{1}{\sqrt{2}}\begin{pmatrix}\sin v & -1 & \cos v\\
\sqrt{2}\cos v & 0 & -\sqrt{2}\sin v\\
\sin v & 1 & \cos v
\end{pmatrix}.
\end{equation}

A generic interaction between the three different helical modes is
described by the following Hamiltonian 
\begin{equation}
H_{{\rm int}}=\sum_{i,a}V_{0}n_{i,a}n_{i,a}+\sum_{i,a\neq b}V_{ab}n_{i,a}n_{i,b},
\end{equation}
where the density at each site $i$ and channel $a$ is $n_{i,a}=\sum_{\sigma}(c_{i,a}^\sigma)^{\dagger}c_{i,a}^{\sigma}.$
This interaction parameters are symmetric $V_{ab}=V_{ba}$.

After bosonization, using the basis (\ref{field_def}) of the main text, the forward scattering Hamiltonian becomes 
\begin{eqnarray}\nonumber
 H_{\rm fs}&=&\sum_{a=c,1,2}\frac{v_{a}}{2}\int dx\left(K_{a}(\partial_{x}\tilde{\varphi}_{2})^{2}+\frac{1}{K_{a}}(\partial_{x}\tilde{\theta}_{2})^{2}\right)\\
&+&\int dx(\zeta_{1}\partial_{x}\tilde{\varphi}_{c}\partial_{x}\tilde{\varphi}_{2}+\zeta_{2}\partial_{x}\tilde{\theta}_{c}\partial_{x}\tilde{\theta}_{2}),
\end{eqnarray}
where the parameters $v_a, K_a, \zeta_{1,2}$ satisfy
\begin{widetext}
\begin{eqnarray}
v_{1} & = & \tilde{v}_{F}-\frac{g_{2}}{4\pi}+\frac{g-g'}{2\pi},\quad K_1=1,\quad \zeta_{1} = \frac{\sqrt{2}}{6\pi}(g'-g),\quad \zeta_{2}=\frac{\sqrt{2}}{6\pi}(3g-g_{1}-g')\\
v_{2}K_{2}  &= & \tilde{v}_{F}-\frac{g_{2}}{4\pi}+\frac{g'-g}{6\pi},\quad
v_{2}K_{2}^{-1}  =  \tilde{v}_{F}-\frac{g_{2}}{4\pi}+\frac{g}{2\pi}+\frac{5g'}{6\pi}-\frac{2g_{1}}{3\pi},\\
v_{c}K_{c} & = & \tilde{v}_{F}+\frac{g_{2}}{2\pi}+\frac{g'-g}{3\pi},\quad
v_{c}K_{c}^{-1}  =  \tilde{v}_{F}+\frac{g_{2}}{2\pi}+\frac{2(g'+g_{1})}{3\pi}+\frac{g+3\tilde{V}_{0}}{\pi}.
\end{eqnarray}
In terms of the microscopic parameters, we have the relations $\tilde{V}_0=\frac{a_0}{4}(V_0+V_{12}+V_{13}+V_{23})$ and  
\begin{eqnarray}\label{parameters_gen}
\frac{g}{a_0} & = & \frac{V_0-V_{13}}{8}(1+\cos^22v)+\frac{V_{23}-V_{12}}{4}\cos2v,\quad
\frac{g'}{a_0}  =  \frac{V_0+V_{13}-V_{23}-V_{12}}{4}+\frac{V_0-V_{13}}{2}\cos^{2}2v,\\
\frac{g_{1}}{a_0} &= & \frac{\cos2v}{2}(V_{12}-V_{23}-(V_0-V_{13})\cos2v),\quad
g_{2}  =  2(2g+g_1),\quad 
g_{1}'= 2(g'-g)+g_{2}. \label{parameters_fin}
\end{eqnarray}
\end{widetext}
The complete Hamiltonian reads 
\begin{eqnarray}
H &=&  H_{\rm fs}+\bar{g}'_{1}\int dx\cos(2\sqrt{2\pi}\tilde{\varphi_{1}})\\\nonumber
&+& 2\bar{g}_{2}\int dx\left(\cos(\sqrt{6\pi}\tilde{\varphi}_{2})+\cos(\sqrt{6\pi}\tilde{\theta}_{2})\right)\cos(\sqrt{2\pi}\tilde{\varphi}_{1}). 
\end{eqnarray}
with $\bar{g}=\frac{g}{2(\pi a_{0})^{2}}$. The transition line between the ET and TRSB phases is defined by $g+g'-g_{1}=0$.
The symmetric limit $t_L=t_R$  corresponds to $v=\pi/4$. For this value the general model reduces to the one we used in the main part of the manuscript.
Assuming that $g_1>0$ we can define $\chi=\frac{g+g'}{g_1}$ and use this parameter to characterise the transition between the TRSB and the ET phase.
In this case $\chi>1$ implies that the system flows into the TRSB phase, while $\chi<1$ implies that the system flows to the ET phase. 
The renormalized parameter $[\chi]_{\rm IR}$ obtained from the RG flow of the interaction constants $g,g',g_1$ determines the properties of the low energy theory,
so it takes the role of $h/J$ in the Hamiltonian $H_{\rm eff}$ of Eq. (\ref{H_eff}). A similar analysis can be done in the regime where $g_1<0$. 
For $g_1=0$, the system is always in the TRSB phase if $g+g'>0$, while if $g+g'<0$ the system is always in the ET phase. To obtain a transition in this case, the 
parameters $g, g'$ should have opposite signs. The case $g>0, g'<0$ is discussed in the main text. The opposite case of $g'>0, g<0$ can be obtained from the previous one by 
interchanging the roles of $g$ and $g'$.

\section {Scattering Matrix for non-interacting channels}
\label{scattering_matrix}

The Schr\"{o}dinger equation for three chiral fermions scattering off an impurity at $x=0$ can be written as
\begin{equation}
\label{Schrodinger}
iv_F(1\otimes\sigma_z)\partial_x \Psi +\mathcal{V} \delta(x)\Psi=E\Psi.
\end{equation}
Here $\mathcal{V}$ parameterizes the scatterer and $\Psi$ is a 6-component spinor that contains the right and left mover part of the chiral fermion.
This scatterer potential can be decomposed in the basis $V=\sum_aV_a\otimes\sigma^a$, where $\sigma^a$ are the Pauli and the $2\times 2$ identity matrices.
The matrices $V_a$ act in the channel space, while $\sigma^a$ acts between the chiralities of the fermions.
Without losing generality, the backscattering part of the potential can be written in the form $V_x\otimes \sigma_x$, where TR symmetry dictates that the
scatterer potential $V$  in the is such that $V_x^T=-V_x$. Taking the determinant of this equation, we find that $\det(V_x^T)=(-1)^3\det(V_x)=0$. 
Also follows from the antisymmetry of $V_x$ that its the trace vanishes. In the basis $\tilde{\Psi}=(U_V\otimes 1_{2\times2})\Psi$ that diagonalizes $V_x$ Eq. 
(\ref{Schrodinger}) splits into
\begin{eqnarray}&&
\label{Schrodinger_2}
iv_F(1\otimes\sigma_z)\partial_x \tilde{\Psi} _1+ ir \sigma_x\delta(x)\tilde{\Psi}_1=E\tilde{\Psi}_1\,,\nonumber \\&&
iv_F(1\otimes\sigma_z)\partial_x \tilde{\Psi}_2 =E\tilde{\Psi}_2\,,\nonumber \\&&
iv_F(1\otimes\sigma_z)\partial_x \tilde{\Psi}_3 -ir \sigma_x\delta(x)\tilde{\Psi}_3=E\tilde{\Psi}_3\,.
\end{eqnarray}
where $r$ is one of the eigenvalues of $V_x$ and parameterizes the strength of the scattering potential.
These equations describe the propagation of three decoupled modes, that constitute independent conducting channels.
Due to TR symmetry and the number of channels being odd, there is one mode with zero reflection across the impurity.
Solving the previous equations using the regularisation $\int dx\delta(x)\Psi(0)=\frac{1}{2}(\Psi(0_+)+\Psi(0_-))$, we find the scattering matrix
for the modes 1 and 3 to be 
\begin{equation}
S=\cos\beta\sigma^0\pm i\sin\beta\sigma^x,
\end{equation}
where the $+(-)$ sign is for the mode 1(3). Here $\beta=2\tan^{-1}\frac{r}{2v_F}.$
Defining $g_{\rm imp}\equiv\frac{r}{2v_F}$, it is found \cite{McKellar1987} that the transmission coefficient $T_i$ for mode 1 and 3 is
\begin{equation}
 T_1=T_3=\frac{1-g^2_{\rm imp}}{1+g^2_{\rm imp}}.
\end{equation}

\end{appendix}

\bibliography{biblio.bib}

\end{document}